\documentclass[twocolumn]{aastex701}

\begin{document}

\title{Igniting galaxy formation in the post-reionization universe}

\author[orcid=0000-0002-3430-3232,sname='Moreno']{Jorge Moreno}
\affiliation{Department of Physics and Astronomy, Pomona College, Claremont, CA 91711, USA}
\affiliation{Carnegie Observatories, 813 Santa Barbara St., Pasadena, CA 91101, USA}
\email[show]{jorge.moreno@pomona.edu}  

\author[0000-0002-2651-7281]{Coral Wheeler}
\affiliation{Department of Physics and Astronomy, California State Polytechnic University, Pomona, Pomona, CA 91768, USA}
\email{cwheeler@cpp.edu}

\author[0000-0002-5908-737X]{Francisco~J. Mercado}
\affiliation{Department of Physics and Astronomy, Pomona College, Claremont, CA 91711, USA}
\affiliation{TAPIR, Mailcode 350-17, California Institute of Technology, Pasadena, CA 91125, USA}
\email{francisco.mercado@pomona.edu}

\author[0000-0003-1848-5571]{M.~Katy Rodriguez Wimberly}
\affiliation{Department of Physics and Astronomy, California State University, San Bernardino, San Bernardino, CA 92407, USA}
\email{maria.wimberly@csusb.edu}

\author[0000-0002-8429-4100]{Jenna Samuel}
\affiliation{Department of Astronomy, The University of Texas at Austin, 2515 Speedway, Stop C1400, Austin, TX 78712-1205, USA}
\affiliation{Cosmic Frontier Center, The University of Texas at Austin, Austin, TX 78712}
\email{jenna.samuel@austin.utexas.edu}

\author[0000-0003-0965-605X]{Pratik~J. Gandhi}
\affiliation{Department of Astronomy, Yale University, New Haven, CT 06520, USA}
\email{pratik.gandhi@yale.edu}

\author[0000-0002-0766-1704]{Elia Cenci}
\affiliation{Department of Astrophysics, Universität Zürich, Zurich, CH-8057, Switzerland}
\email{elia.cenci@uzh.ch}

\author[0000-0002-1109-1919]{Robert Feldmann}
\affiliation{Department of Astrophysics, Universität Zürich, Zurich, CH-8057, Switzerland}
\email{feldmann@physik.uzh.ch}

\author[0000-0002-9604-343X]{Michael Boylan-Kolchin}
\affiliation{Department of Astronomy, The University of Texas at Austin, 2515 Speedway, Stop C1400, Austin, TX 78712-1205, USA}
\affiliation{Cosmic Frontier Center, The University of Texas at Austin, Austin, TX 78712}
\email{mbk@astro.as.utexas.edu}

\author[0000-0003-0603-8942]{Andrew Wetzel}
\affiliation{Department of Physics \& Astronomy, University of California, Davis, CA 95616, USA}
\email{awetzel@ucdavis.edu}

\author[0000-0003-4298-5082]{James~S. Bullock}
\affiliation{Department of Physics and Astronomy, University of Southern California, Los Angeles, CA 90089, USA}
\email{dean@dornsife.usc.edu}

\author[0000-0003-3729-1684]{Philip~F. Hopkins}
\affiliation{TAPIR, Mailcode 350-17, California Institute of Technology, Pasadena, CA 91125, USA}
\email{phopkins@caltech.edu}

\begin{abstract}

It is widely believed that the ultraviolet background produced during the epoch of reionization conspires against the formation of low-mass galaxies. Indeed, this mechanism is often invoked as part of the solution to the so-called `missing satellites problem.' In this paper we employ \texttt{FIREbox}, a large-volume cosmological simulation based on the Feedback In Realistic Environments (\texttt{FIRE-2}) physics model, to characterize the mechanisms governing galaxy ignition in the post-reionization era. By carefully matching recently-ignited halos (with stellar ages below $100$ Myr at the time of selection) to halos that failed to form any stars, we conclude that the presence of cold-dense gas and halo concentration help incite the process of galaxy formation. Concretely, we find that $100\%$ of recently-ignited halos experience cold-dense gas enhancements relative to their matched failed counterparts. Likewise, approximately $83\%$ display enhancements in both cold-dense gas and Navarro-Frenk-White concentration ($c_{\rm NFW}$), while the remaining $\sim17\%$ exhibit enhanced cold-dense gas content and suppressed $c_{\rm NFW}$ values. Lastly, our simulation suggests that galaxy ignition can occur as late as $z=2$, potentially allowing us to observationally catch this process `in the act' in the { foreseeable} future.

\end{abstract}

\keywords{galaxies --- formation --- galaxies --- star formation --- galaxies --- ISM --- galaxies --- halos --- cosmology --- large-scale structure of universe}

\section{Introduction}\label{sec:intro}

It has been 47 years since \cite{WhiteRees1978} first proposed that galaxies form from the condensation of gas at the centers of dark matter halos. Eight years later, \cite{Rees1986} put forward the idea that the cosmic ultraviolet (UV) background radiation produced during the epoch of reionization is able to suppress the formation of low-mass\footnote{We humbly ask the astronomical community to refrain from using the term `dwarf' to describe low-mass galaxies, as this usage is derogatory and exclusionary against people with dwarfism and adds no scientific value.} galaxies \citep[see also][]{BabulRees1992,Efstathiou1992,Shapiro1994,ThoulWeinberg1996,Quinn1996,BarkanaLoeb1999}. Indeed, this mechanism has been instrumental in helping solve the mismatch between the observed number of galaxies in the Local Group and predictions from cosmological dark matter only simulations -- the so-called `missing satellites problem' \citep{Klypin1999,Moore1999,Bullock2000,Benson2002,Somerville2002}.

\begin{figure*}
    \centering
    \includegraphics[width=0.329\textwidth]{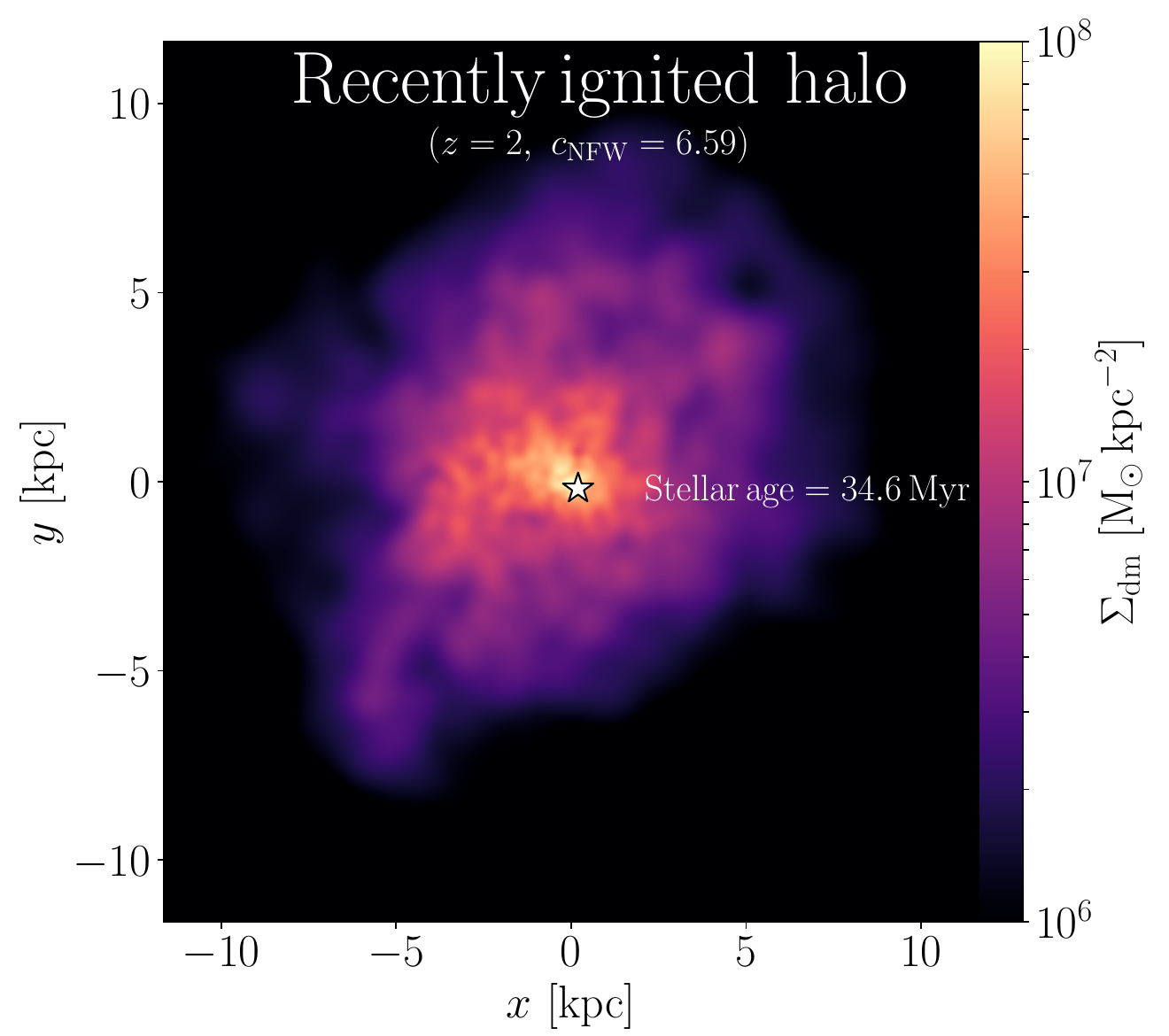}
    \includegraphics[width=0.329\textwidth]{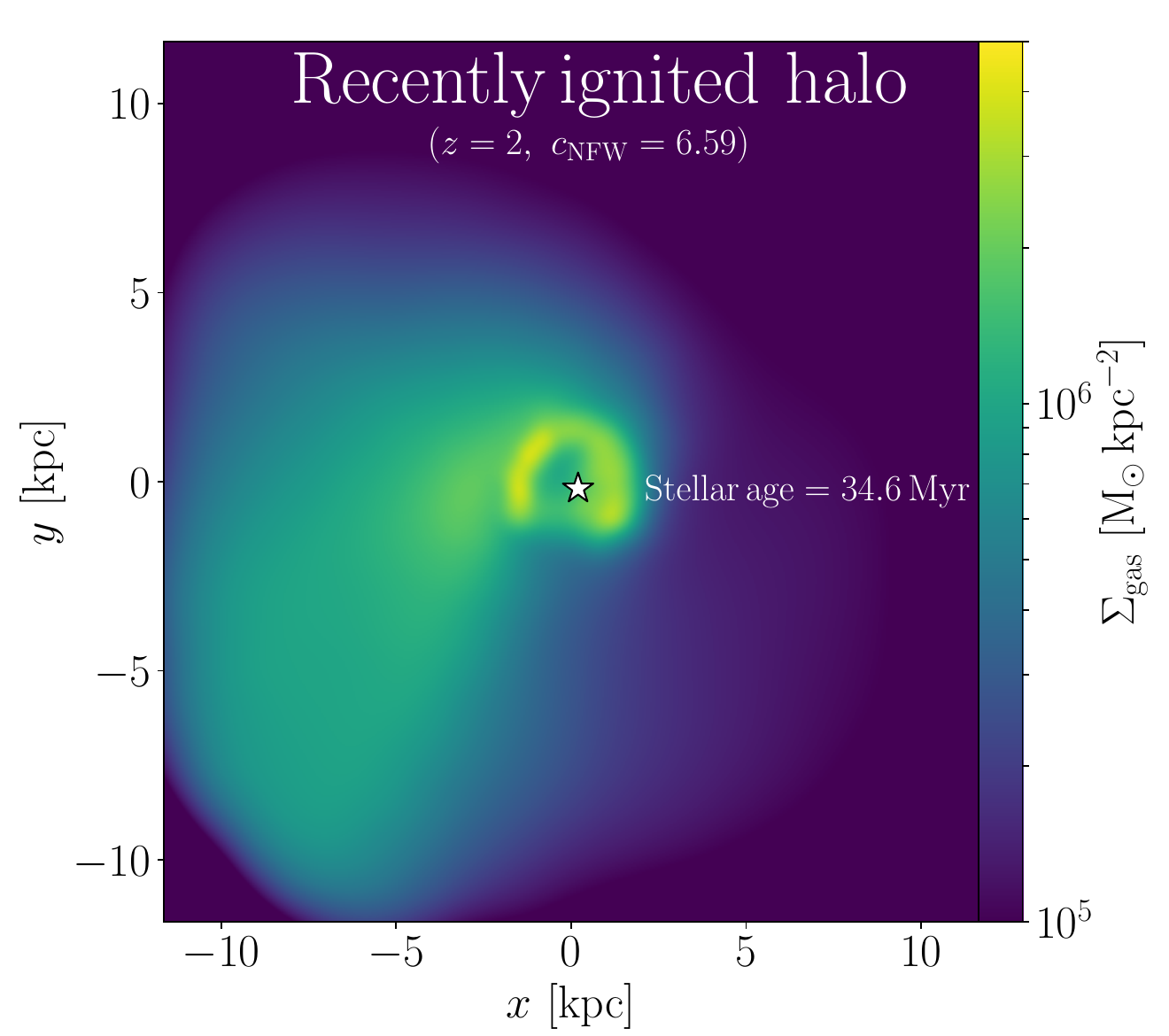}
    \includegraphics[width=0.329\textwidth]{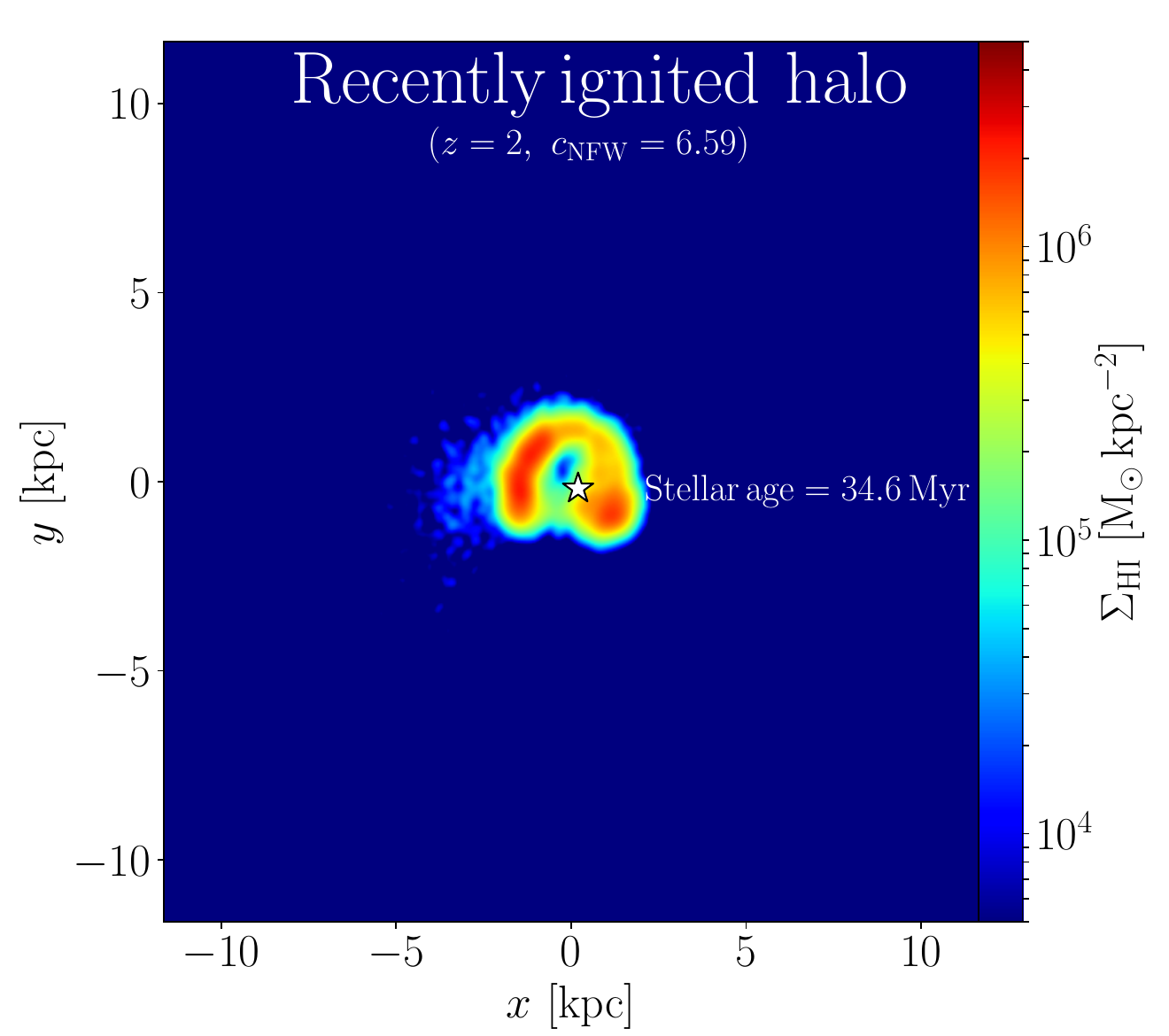}
    \includegraphics[width=0.329\textwidth]{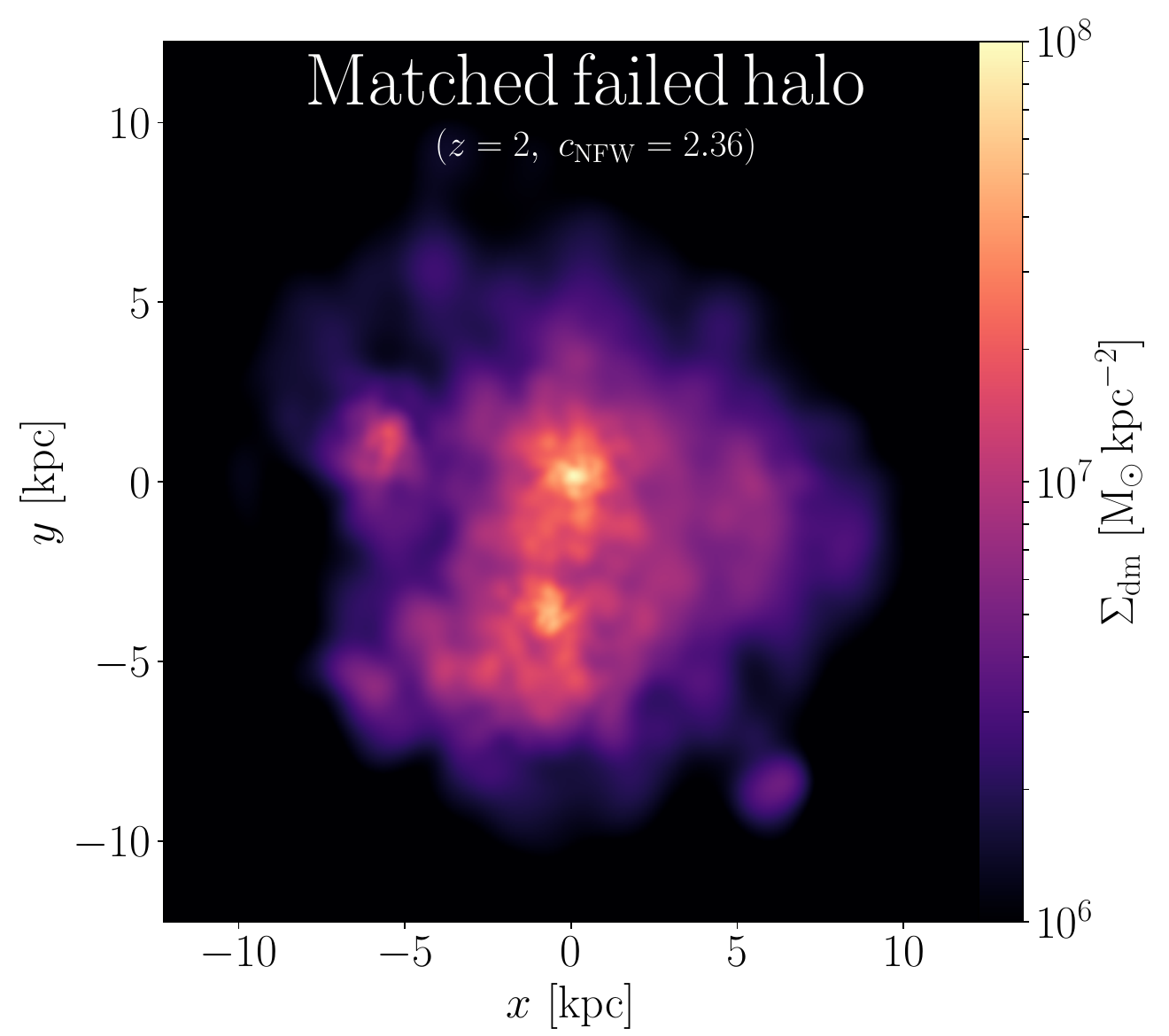}
    \includegraphics[width=0.329\textwidth]{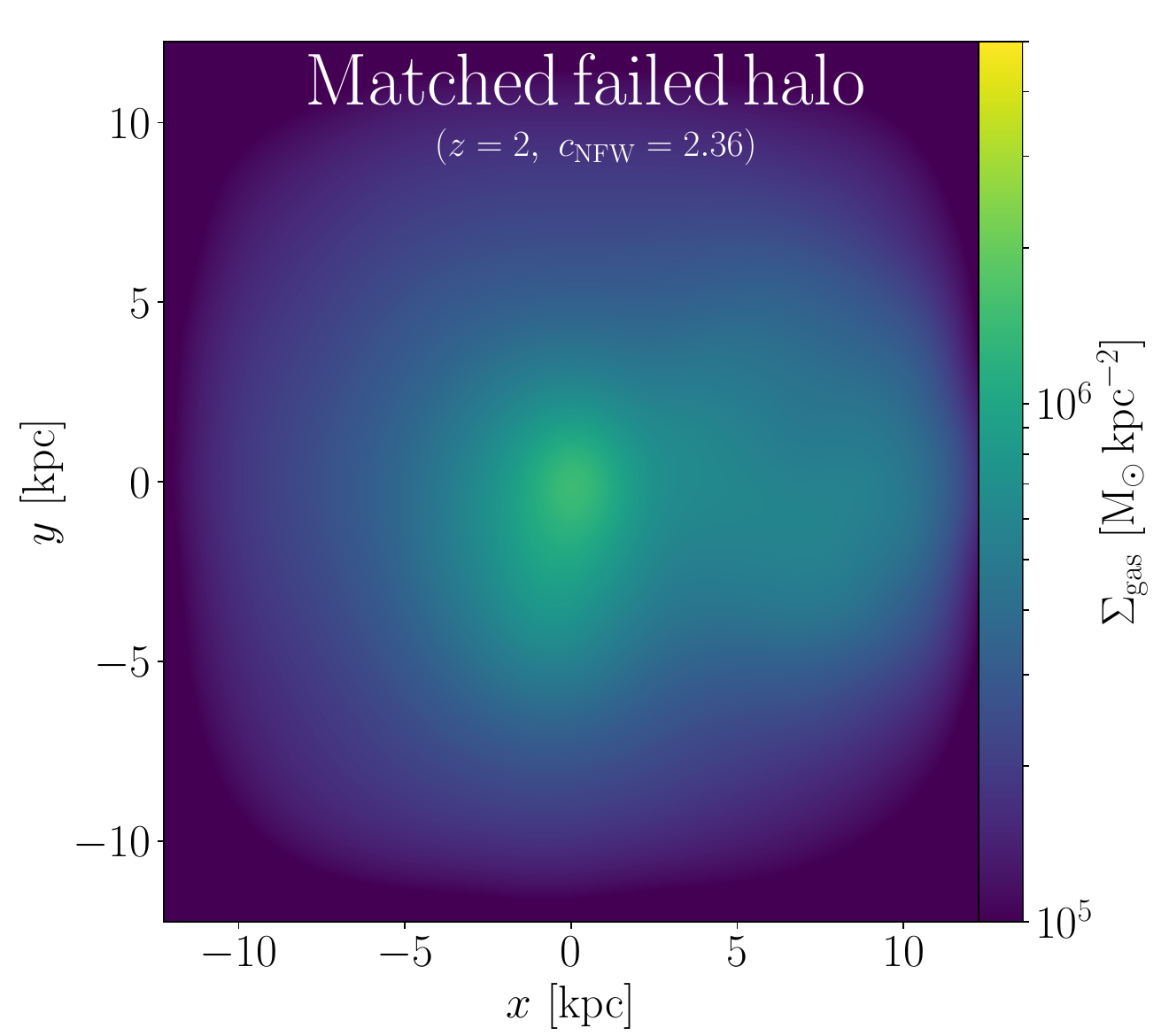}
    \includegraphics[width=0.329\textwidth]{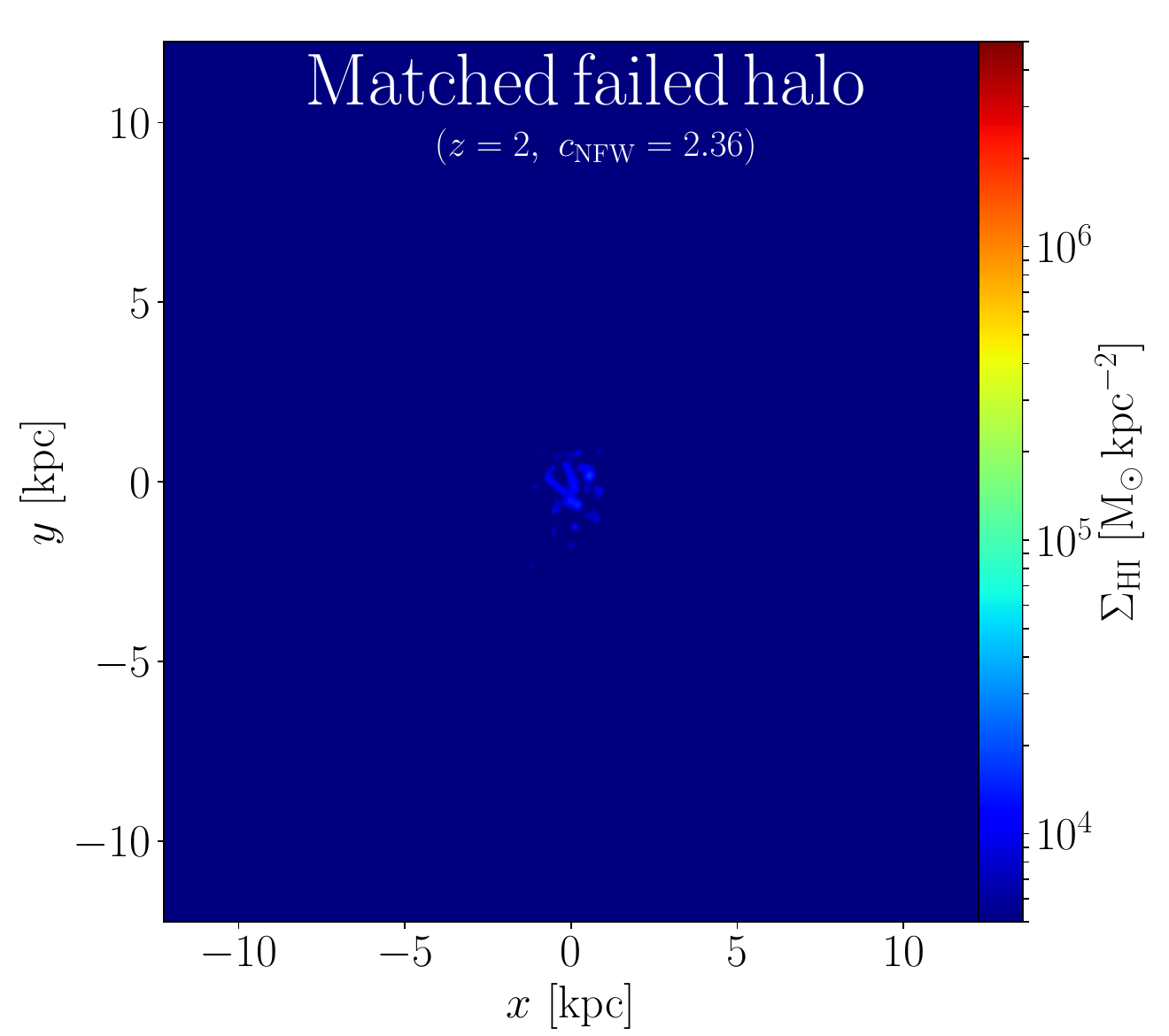}
    \caption{Archetypal recently-ignited and (matched) failed halos in our simulation. The panels show two halos at $z=2$ matched to have nearly identical virial and gas masses ($M_{\rm vir} \sim 1.29 \times 10^9 M_\odot$ and $M_{\rm gas} \sim 1.17 \times 10^8 M_\odot$). The field of view encompasses the virial radius. {\it Top panels:} Recently-ignited halo with a single star particle, denoted as a white star-shaped symbol, with stellar mass $M_\star = 6.26 \times 10^4 M_\odot$ and stellar age $t^{\star}_{\rm age}= 34.6$ Myr. {\it Bottom panels:} Matched failed halo. {\it Left-to-right panels:} Dark matter, gas and HI gas surface mass density maps. Although matched in $M_{\rm vir}$, $M_{\rm gas}$ and $z$, the recently-ignited halo is more concentrated than its failed counterpart ($c_{\rm NFW} = 6.59$ versus $2.36$). The ignited halo exhibits a dense, complex horseshoe-shaped central gas concentration, particularly salient in the HI component. The gas in the matched failed halo is extended and diffuse, with a barely detectable central HI contribution. This figure shows the only ignited halo in our sample found at $z=2$ with stellar age below $100$ Myr, with no other such halos identified at lower redshifts.
    }
    \label{fig:awake_vs_failed}
\end{figure*}

But how does this suppression process really unfold in detail? Here, the ionizing UV background (UVB) interjects against the formation of low-mass galaxies on several fronts: by heating up the intergalactic medium (IGM), which can either (1) inhibit accretion \citep{NohMcQuinn2014,Jeon2025} or (2) pre-process infalling gas \citep{ThoulWeinberg1996}; (3) by suppressing cooling of gas already captured by halos \citep{Wise2007}; or (4) by evacuating halos of their gas via photo-evaporation \citep{BarkanaLoeb1999,Okamoto2008uv,Nickeson2011}. To paraphrase, the complex interplay between the ionizing UVB and gravity dictates a halo's gas endowment and whether or not said fuel is ultimately eligible for investment into new stars.

Many authors attempt to encapsulate this complex pipeline into simple analytic schemes in which a critical mass (or circular velocity) scale governs the onset of galaxy formation. These include the balance between photoheating and cooling \citep{Rees1986}, Jeans-mass arguments \citep{Haiman1996}, scales over which baryonic fluctuations are smoothed in linear perturbation theory \citep{Gnedin2000}, and redshift-dependent critical mass scales \citep{Benitez-Llambay2020} controlled by atomic cooling (pre-reionization) or the ability for gravity and the pressure of the IGM to maintain gas in hydrostatic equilibrium (post-reionization). 

There is no doubt that these schemes are informative. They do provide insight on the net effect of reionization on the suppression of galaxy formation at the smallest scales, and supply the seeds for the first galaxies in semi-analytic models \citep{Benson2002,Somerville2002,Ahvazi2024}. Nevertheless, these models generally fail to quantify the relative importance and relevant scales of the various steps in the aforementioned multi-layered pipeline. 

Our approach in this paper (and forthcoming work) is to examine this pipeline one step at a time. One could ask which halos fail to accrete their fair share of baryons and why? Or which halos fail to retain their gas and why? As a first step, however, we elect to focus on the survivors, those halos which were able to retain their gas, and which may or may not be able to form stars. At this stage, we also avoid cosmic times in which the reionization of the universe is still unfolding, and elect to focus on the post-reionization era. 

Concretely, this paper asks the following pointed question: {\it what conditions must a halo with a resolved gas component meet in order to successfully ignite galaxy formation in the post-reionization era?} To answer this, we employ \texttt{FIREbox} \citep{Feldmann2022}, a simulation capable of capturing the multi-phase structure of the interstellar medium (ISM) and feedback-regulated star formation within a cosmologically representative volume. Indeed, many recent works addressing similar questions tend to focus exclusively on either small zoom volumes containing hand-picked isolated low-mass halos \citep{Wheeler2015,Fitts2017,Wheeler2019,Munshi2021,Kim2024}, or halos in the vicinity of Milky Way or Local Group analogs \citep{Sawala2016,Jeon2025}. Nevertheless, in order to quantify the global importance of each process responsible for galaxy ignition across the universe, it is imperative to employ a large and representative volume. 

Indeed, \texttt{FIREbox} is ideal for accomplishing this task. To illustrate, Figure~\ref{fig:awake_vs_failed} shows a recently ignited halo at $z=2$ (top panels). Identified as the only object of its kind at this redshift, this object hosts a stellar population formed $34.6$ Myr prior (white star-shaped symbol). This is the latest epoch in the history of the universe exhibiting such a halo in our simulation. This particular object has virial mass $M_{\rm vir} = 1.29 \times 10^9 M_{\odot}$, gas mass $M_{\rm gas} = 1.17 \times 10^{8} M_\odot$ and Navarro-Frenk-White (NFW) concentration $c_{\rm NFW}=6.59$ \citep{NFW}. Its gas content consists of an extended halo and a prominent horseshoe-shaped central concentration containing $M_{\rm HI} = 4.94 \times 10^6 M_\odot$ in neutral gas. For comparison, we also include a failed halo, matched in $M_{\rm vir}$, $M_{\rm gas}$ and $z$ (bottom panels). Compared to its ignited counterpart, this object has a lower NFW concentration ($c_{\rm NFW}=2.36$; in this case, possibly caused by the presence of a massive subhalo) and a substantially smaller neutral gas central reservoir ($M_{\rm HI} = 5.75 \times 10^4 M_\odot$) -- i.e., it appears to not have the necessary fuel for star formation { (although the presence of a secondary massive subhalo suggest that this system could eventually gain a higher concentration and colder/denser ISM ingredients, making it eligible for galaxy ignition in the not-so-distant future).} In sum, this example, and our results below, suggest that the properties of the interstellar medium (ISM) and the structure of the host dark matter halo play a critical role in igniting galaxy formation.

This manuscript is organized as follows. Section~\ref{sec:methods} introduces our physics model and our cosmological simulation. Section~\ref{sec:results} presents our results, which include abundances, evolution across epochs, ISM properties and halo structure. In Section~\ref{sec:discussion} we compare our findings against other works and in Section~\ref{sec:conclusions} we list our conclusions and offer future directions.


\section{Methods}\label{sec:methods}

\subsection{The \texttt{FIRE-2} physics model}\label{subsec:fire2}

We employ the Feedback In Realistic Environments\footnote{For information on the \small{FIRE} Project, visit \url{https://fire.northwestern.edu}.} (\texttt{FIRE-2}) model \citep{FIRE2}, which relies on the Meshless Finite-Mass (MFM) version of the \texttt{GIZMO} hydrodynamics code \citep{GIZMO}. This framework has been extensively validated in several publications analyzing properties of galaxies across a wide range of stellar masses and at many numerical resolutions \citep{CAFG2018}. We follow eleven separately-tracked species: H, He, C, N, O, Ne, Mg, Si, S, Ca, and F. Our treatment of radiative heating and cooling includes free-free, photo-ionization/recombination, Compton, photoelectric, dust-collisional, cosmic ray, molecular, metal-line and fine-structure processes. The model accounts for a spatially uniform, meta-galactic UV/X-ray background \citep{CAFG2009} and locally-driven photo-heating and self-shielding. We constrain star formation to Jeans-unstable, self-gravitating (at the resolution scale), self-shielded molecular gas. Stellar feedback mechanisms include momentum flux from radiation pressure; energy, momentum, mass and metal injection from SNe (Types Ia and II); stellar mass loss (OB and AGB stars); plus photoionization and photo-electric heating channels. We calculate core-collapse supernovae rates using \texttt{STARBURST99}, a stellar synthesis model \citep{Leitherer1999}, stellar wind yields from \cite{WiersmaCooling}, Type Ia supernova rates from \cite{Mannucci2006} and yields from \cite{Iwamoto1999}. We assume a \cite{kroupa2001} initial mass function (IMF). 


\begin{figure*}
    \centering
    \includegraphics[width=0.497\textwidth]{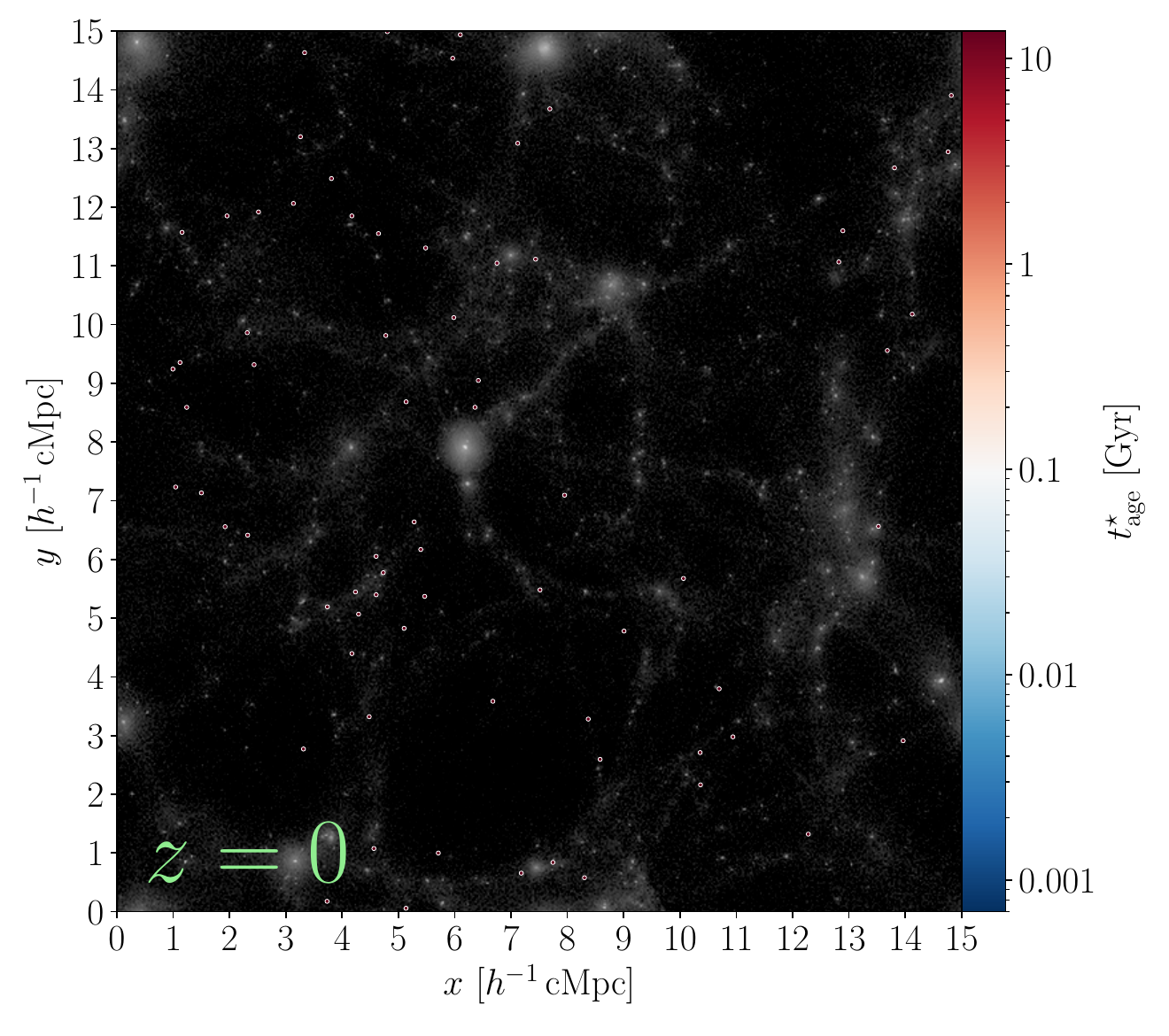}
    \includegraphics[width=0.497\textwidth]{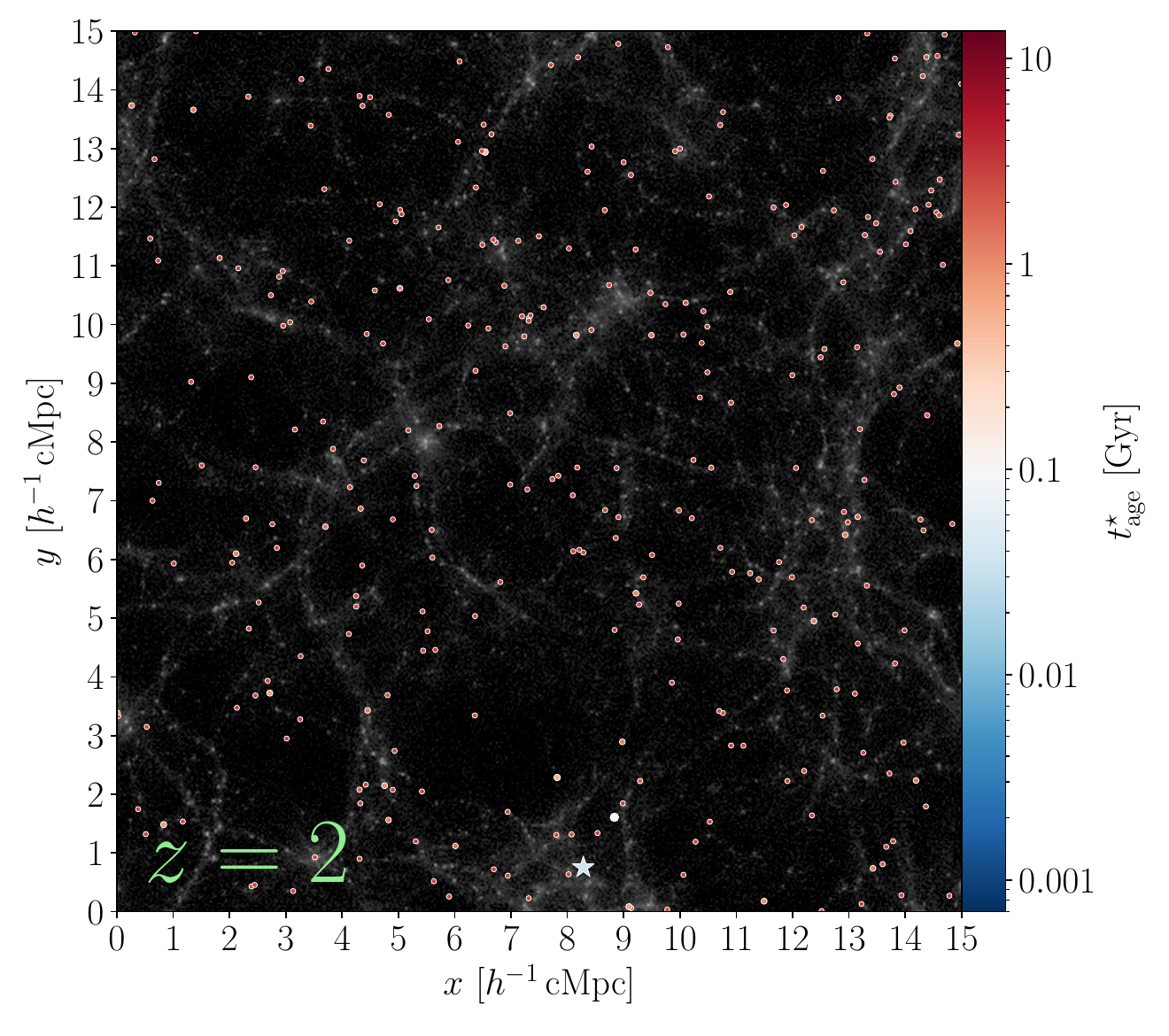}
    \includegraphics[width=0.497\textwidth]{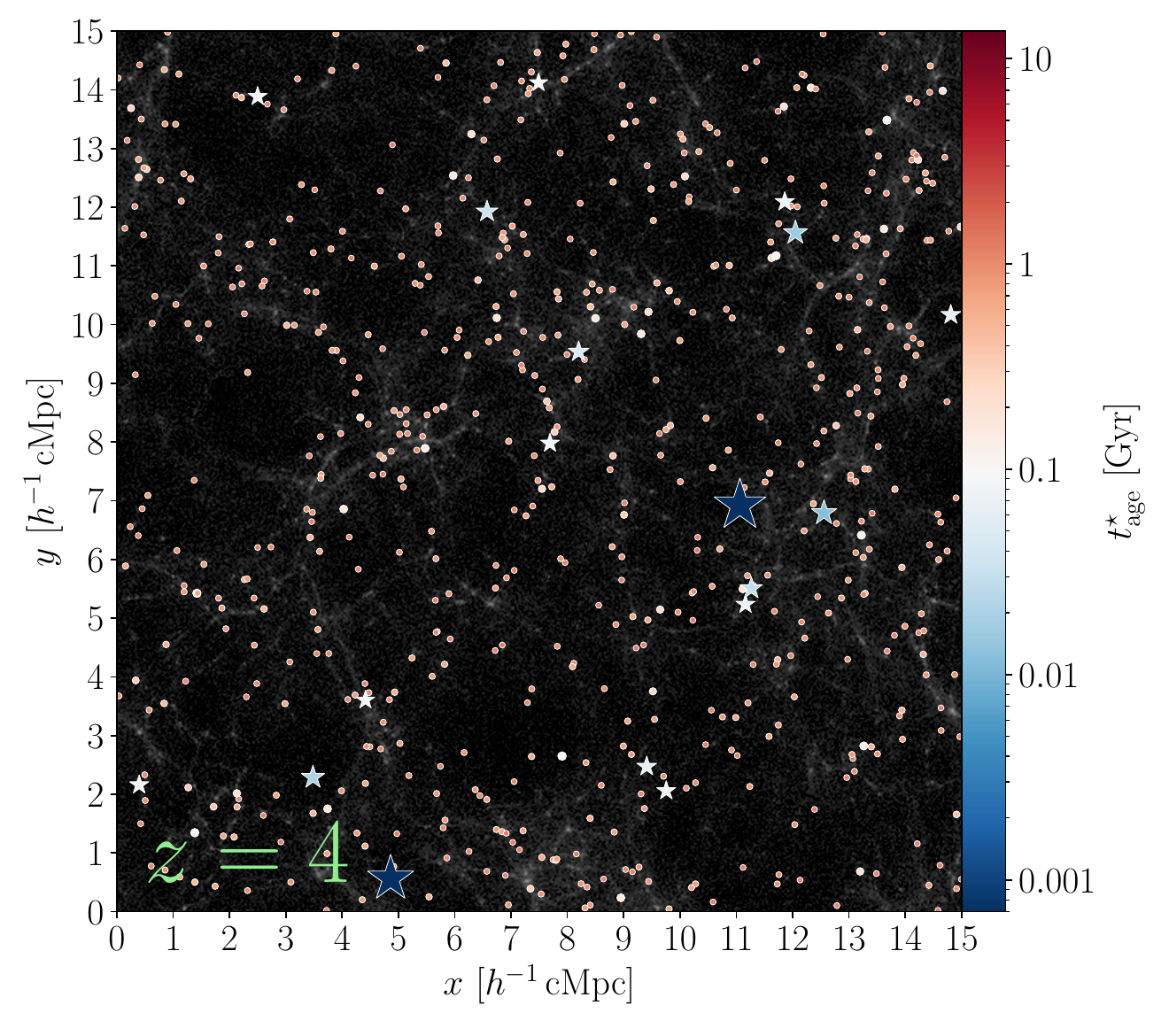}
    \includegraphics[width=0.497\textwidth]{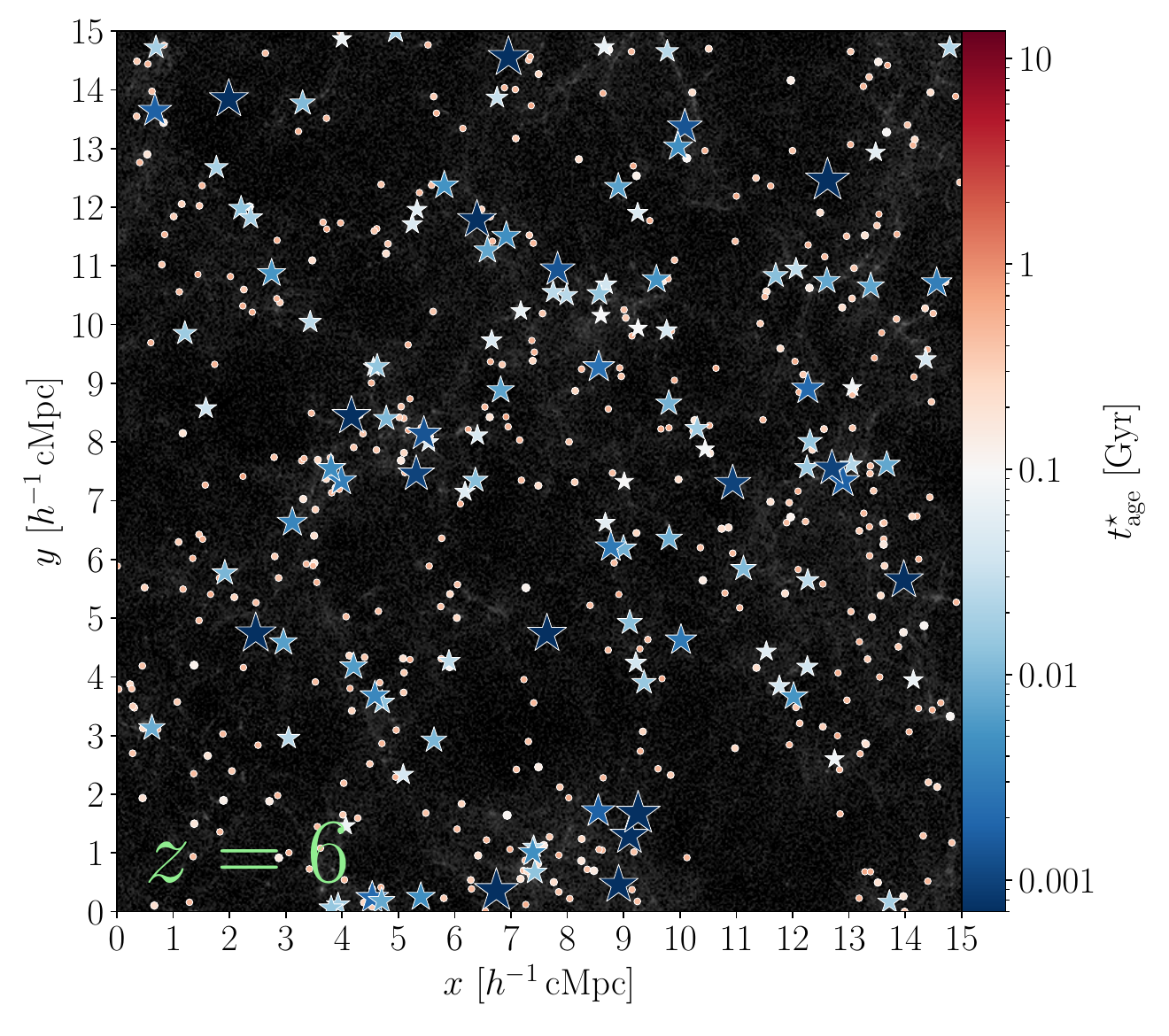}
    \caption{Ignited halos on the cosmic web across epochs. {\it Top-left-to-bottom-right:}  $z=0, 2, 4$ and $6$, indicated by the large green annotations. On the background, each panel shows a surface-density map of our comoving simulation volume. The white symbols represent the location of ignited halos, color coded by stellar age, with the palette centered at $t^{\star}_{\rm age} = 100$ Myr. Symbol sizes increase with decreasing stellar age to emphasize recently ignited halos, which become more numerous with increasing $z$. We also differentiate objects with stellar ages below $100$ Myr from the rest by switching from circular to star-shaped symbols.
    }
    \label{fig:cosmic_web}
\end{figure*}

\subsection{FIREbox}\label{subsec:FIREbox}

\texttt{FIREbox} is one of the few large-volume cosmological simulations capable of resolving the multi-phase turbulent structure of the interstellar medium (ISM) at high resolution \citep{Feldmann2022}. With a periodic box of $22.1\, {\rm cMpc}\,$ on the side, it is large enough to capture the evolution of galaxies and halos in diverse environments, and is ideal for generating and characterizing the abundance and nature of rare objects in a cosmologically meaningful way \citep{Moreno2022}. The simulation has the following number of baryonic and dark matter particles: $N_{\rm b}=1024^3$ and $N_{\rm dm}=1024^3$. Baryonic particle masses are initially set to $m_{\rm b}=6.3\times10^4\,M_{\odot}$ for gas and star particles, while the dark matter particles have $m_{\rm dm}=3.3\times10^5\, M_{\odot}$. The force resolution is set at a fixed $h_{\rm dm}=80$ pc (physical) for dark matter  particles and 12 pc for star particles. Force resolution for gas is set to equal the adaptive smoothing length down to a minimum of 1.5 pc, which occurs only in the densest regions of galaxies. We identify galaxies and halos -- with or without galaxies -- using the \texttt{AMIGA Halo Finder} \citep[\texttt{AHF},][]{Knollmann2009}, which employs an iterative unbinding procedure to identify gravitationally-bound objects \citep{Knebe2011}, and assumes the virial-mass definition of \cite{Bryan1998} by design. We use \texttt{yt} \citep{Turk2011} to interface with the particle data. We adopt the following cosmological parameters: $\Omega_{\rm m} = 0.3089$, $\Omega_{\Lambda} = 0.6911$, $\Omega_{\rm b} = 0.0486$, $\sigma_{8} = 0.8159$ and $h = 0.6774$.

\begin{deluxetable*}{rl}
\digitalasset
\tablewidth{0pt}
\tablecaption{Halo sample definitions \label{table:samples}}
\tablehead{
\colhead{Sample name} & \colhead{Definition}
}
\startdata
all halos & centrals with $N_{\rm dm} \geq 128$ \\
all failed halos & centrals with $N_{\rm dm} \geq 128$ and $N_{\star} = 0$ \\
all ignited halos & centrals with $N_{\rm dm} \geq 128$ and $N_{\star} = 1$ \\
gaseous halos & centrals with $N_{\rm dm} \geq 128$ and $N_{\rm gas} \geq 128$\\
failed (gaseous) halos & centrals with $N_{\rm dm} \geq 128$, $N_{\rm gas} \geq 128$ and $N_{\star} = 0$ \\
ignited (gaseous) halos & centrals with $N_{\rm dm} \geq 128$, $N_{\rm gas} \geq 128$ and $N_{\star} = 1$ \\
recently-ignited (gaseous) halos & centrals with $N_{\rm dm} \geq 128$, $N_{\rm gas} \geq 128$, $N_{\star} = 1$ and $t^{\star}_{\rm age} \leq 100$ Myr\\
\enddata
\tablecomments{Here, $N_{\rm dm}$, $N_{\rm gas}$ and $N_{\star}$ denote dark-matter, gas and stellar particle numbers respectively, while $t^{\star}_{\rm age}$ represents the stellar-age of the star particle in halos with $N_{\star}=1$. We drop the `central' adjective because we exclude subhalos entirely from this analysis. We also drop the qualifier `gaseous' from the above subsamples (where this word is indicated in parentheses) because this paper focuses primarily on objects with a resolved gaseous component (i.e., with $N_{\rm gas} \geq 128$).}
\end{deluxetable*}


\subsection{Our halo samples}\label{subsec:samples}

In this work we focus exclusively on central halos (throughout this paper, we occasionally use the words `halo' and `central' interchangeably). This is to minimize the role of environment in creating halos with unusual stellar-to-total mass ratios \citep{Penarrubia2008,Errani2021,Jackson2021,Moreno2022,Errani2024,Errani2024u1}. In every case, we require halos to have at least 128 dark matter particles\footnote{Particle number thresholds near 100 are often employed because this is where the Stirling approximation is accurate within 1\%. We adopt 128 here simply because it is a power of 2.}. Hereafter we refer to this set as the {\bf `all halo'} population. We also define two subsets: (1) {\bf `all failed halo'} sample, with zero stellar particles ($N_\star = 0$ -- i.e., these objects have never formed a star particle); and (2) {\bf `gaseous halos'}, with at least 128 gas particles ($N_{\rm gas} \geq 128$). The intersection of these two subsets constitutes the {\bf `failed gaseous halo'} population (hereafter we drop the word `gaseous' and instead use the word `all' to distinguish the parent population from this subset).

We define the {\bf `all ignited halo'} population as that including those halos with only one stellar particle ($N_\star=1$, we elaborate on this choice below). Here, we only consider stellar particles within $0.15R_{\rm vir}$ of the host halo, which corresponds to the typical galactic coverage for more massive systems. This threshold, comparable to the one adopted by \cite{Wheeler2025} -- though slightly larger than the one in \cite{Kravtsov2013} -- helps us avoid the inclusion of star particles formed elsewhere. Namely, it mitigates contamination by stars from nearby systems or accreted halos. From this `all ignited' set, we denote the subset with a resolved gaseous component ($N_{\rm gas} \geq 128$) as the {\bf `ignited halo'} population (we drop the `gaseous' descriptor and use the word `all' to describe the parent sample). 

Note that we elect to focus only on systems with a resolved gas component in order to learn about which characteristics ignite galaxy formation. In fact, this paper focuses primarily on a subsample, the {\bf recently ignited halos}, with stellar age $t^{\star}_{\rm age} < 100$ Myr. We adopt this limit to capture the longer timescales employed to estimate star formation \citep{Flores2020}. Although we do identify systems with ages as short as $1$ Myr, we choose this longer cut as a compromise to guarantee a reasonable sample size without allowing the halo to evolve significantly from its time of ignition. 

We warn the reader against trying to deduce too much from a single stellar-particle halo population. Namely, a single particle contains very limited information, and therefore our goal here is to merely use it as a signpost that the conditions leading to the ignition of galaxy formation been met successfully within our model assumptions. Namely, our aim is to identify halos that have successfully (or not) ignited the process of galaxy formation -- in accordance to our simulation -- by the redshift at which they are identified. Beyond that epoch, these halos may continue forming stars (or begin forming stars in the case of failed objects) at later cosmic times (or not).  

Other authors prefer to utilize stellar-mass, stellar-to-total mass ratio or surface-brightness thresholds to pinpoint halos that successfully formed a galaxy \citep{Lee2024,Jeon2025,Doppel2025}. This paper concentrates primarily on the `recently-ignited' halo population, which is meant to represent halos caught as close as possible to onset of galaxy formation. Allowing for more than one particle runs the risk of selecting halos that have undergone rejuvenation \citep{Ledinauskas2018,Wright2019,Rey2020}, rather than first-time ignition (our primary goal). 

Lastly, we make no claim of having achieved the threshold of galaxy formation with our simulation \citep[][but see \citet{Munshi2021}]{Sawala2016}. Rather, we aim to quantify the frequency of galaxy ignition within a cosmologically-representative volume -- one that employs a physics model that captures the multi-phase structure of the ISM -- and seek connections between the structure and content of halos and their ability to form stars (or not). See Section~\ref{sec:discussion} for a more detailed discussion of the advantages and limitations of our simulations against other works.

For the recently-ignited population, we also construct a {\bf control sample} of gaseous failed halos, matched in $M_{\rm vir}$ and $M_{\rm gas}$ (within $0.1$ dex), plus redshift. We select this tolerance threshold to be smaller than the typical ranges on $M_{\rm vir}$ and $M_{\rm gas}$ for the populations studied in this paper, but large enough to guarantee that each target recently-ignited halo has at least one matched-control failed object. We control on $M_{\rm gas}$ as well because recall that, at this stage, we are interested in learning why a gaseous halo succeeds or fails to make a galaxy -- i.e., without lack of gas being an excuse (we defer investigations on gas-deprived halos to future work). Furthermore, we are also interested in characterizing the ISM in these objects (Section~\ref{subsec:gas}), for which having a resolved gas component is imperative.

In this paper, we report average quantities among each matched control set unless stated otherwise. To compare target halos against their matched control sets, we define the term {\bf enhancement} to represent the ratio of the value of a given property $x$ to the average value of that quantity, $\langle x^{\rm control}\rangle$, among its set of matched controls. We use the term {\bf suppression} to refer to enhancement values below unity. Hereafter, we only report enhancement/suppression values for situations where both $x\neq0$ and $\langle x^{\rm control} \rangle\neq0$.  

This paper focuses only on the post-reionization universe, targeting halos at the following  redshifts: $z=0, 0.5, 1, 2, 3, 4, 5$ and $6$. Our intention is to ask if galaxy formation can be ignited at these later epochs, and what this can teach us about how and why, at a given moment in time, some gaseous halos are ignited or not.
We elect to focus on the post-reionization era in this paper for the following reasons. First, in our attempt to disentangle the complex and multi-layered low-mass galaxy suppression process described in Section~\ref{sec:intro}, it is preferable for the sake of simplicity to focus on epochs where reionization the universe is no longer unfolding. Indeed, past work suggests that the galaxy formation and evolution ought to be treated differently before and after re-ionization \citep{Benitez-Llambay2020,Kim2024}. Therefore, we opt to focus exclusively on the post-reionization era here and defer investigations at earlier epochs to future work (with \texttt{BonFIRE}, a new large-box simulation similar to \texttt{FIREbox}, but with higher-resolution, larger volume, FIRE-3 physics, and targeting the high-redshift universe; Samuel et al., in prep).

Figure~\ref{fig:cosmic_web} shows the evolution of the \texttt{FIREbox} volume at four of these redshifts. The background cosmic web shows the distribution of gas, with pixel brightness indicating mass surface density on a logarithmic scale. The filled white symbols denote the location of ignited halos, color-coded by stellar age. We designate the palette to display the recently-ignited halos (with $t^{\star}_{\rm age} < 100$ Myr) in blue hues. We also adopt symbol sizes that scale inversely with $t^{\star}_{\rm age}$, and switch from circular to star-shaped symbols, to emphasize recently-ignited objects. 

Overall, the number of recently ignited halos increases with redshift. At low redshifts, only objects with old stellar populations exist -- with recently-ignited halos disappearing by $z=2$: pale blue star-shaped symbol at $(x, y) \simeq (8.3,0.8) \, h^{-1}$cMpc (top-right panel). At higher redshifts, this population is more ubiquitous (large star-shaped symbols in dark-blue hues in bottom two panels). We note that the `failed' and `ignited' designations are not generally permanent. A failed halo can always meet the conditions for galaxy formation at a later epoch, and an `ignited' can always experience subsequent star formation -- and either type of halo can also be absorbed by a larger system at later cosmic times. In other words, one should not expect the number of halos indicated by circular or star-shaped symbols Figure~\ref{fig:cosmic_web} to be conserved nor that pairs of symbols can be matched to each other across snapshots. We quantify these populations in the next section.



\begin{figure*}
    \centering
    \vspace{-0.0cm}
    \includegraphics[width=0.325\textwidth]{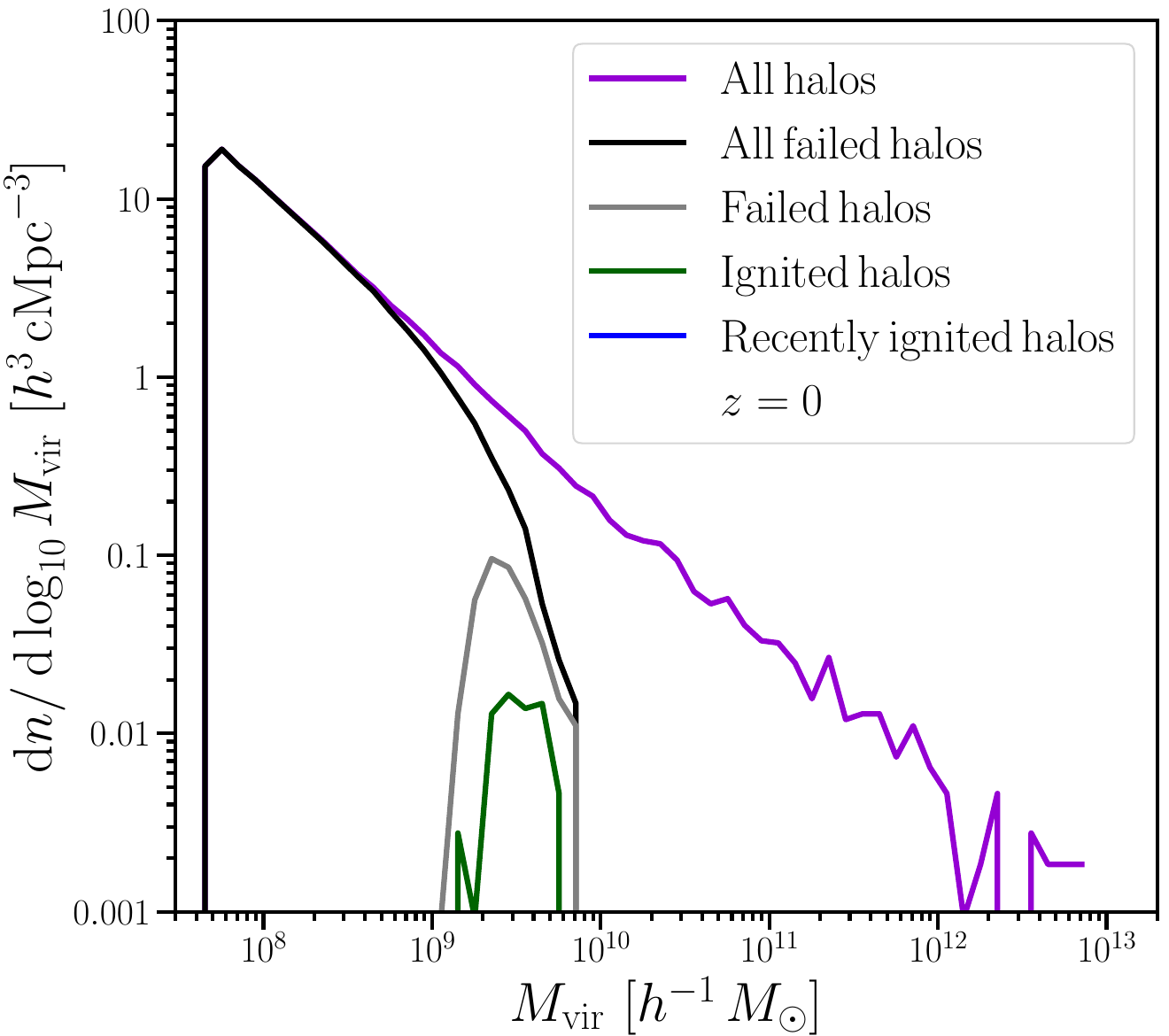}
    \includegraphics[width=0.325\textwidth]{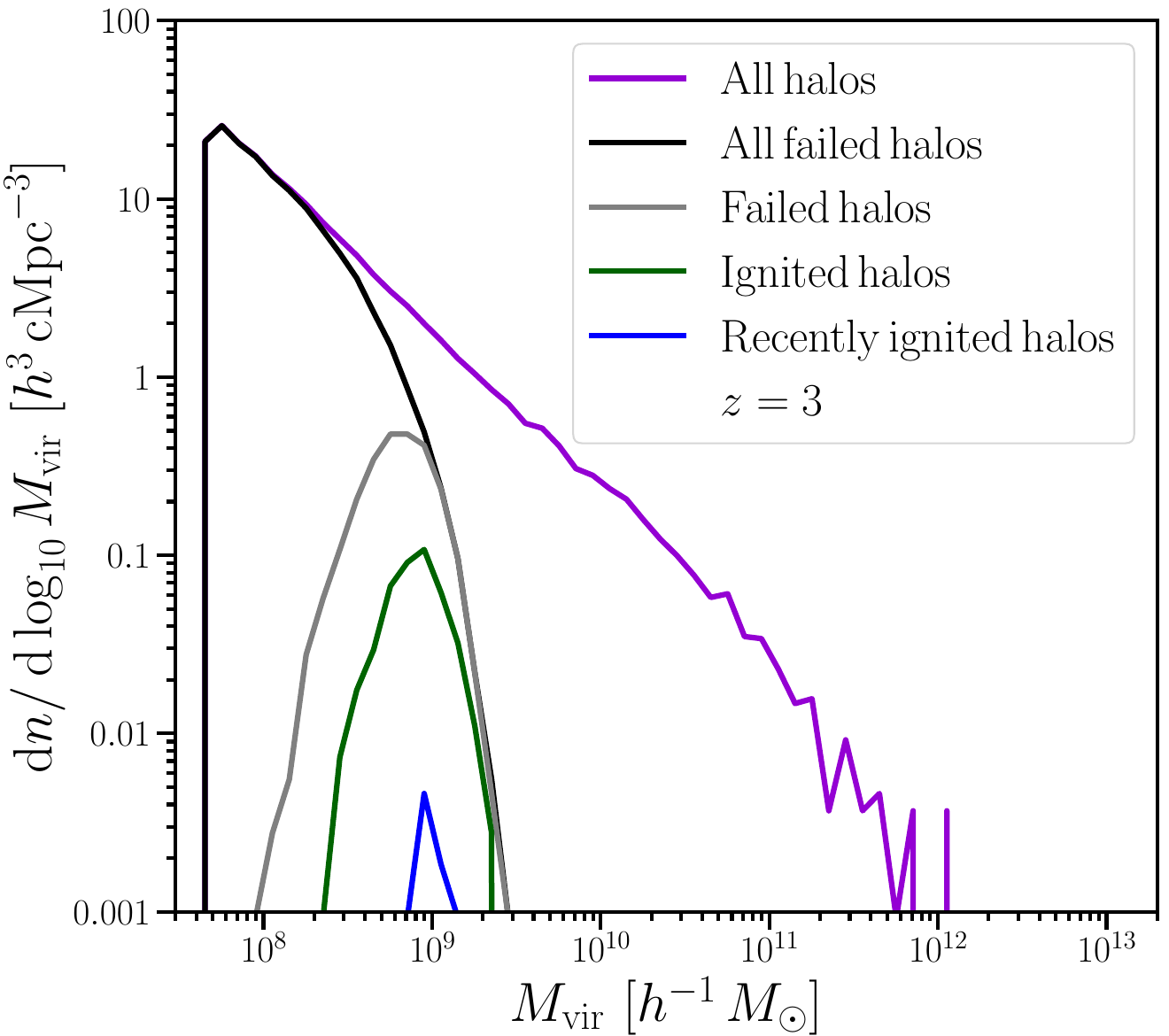}
    \includegraphics[width=0.325\textwidth]{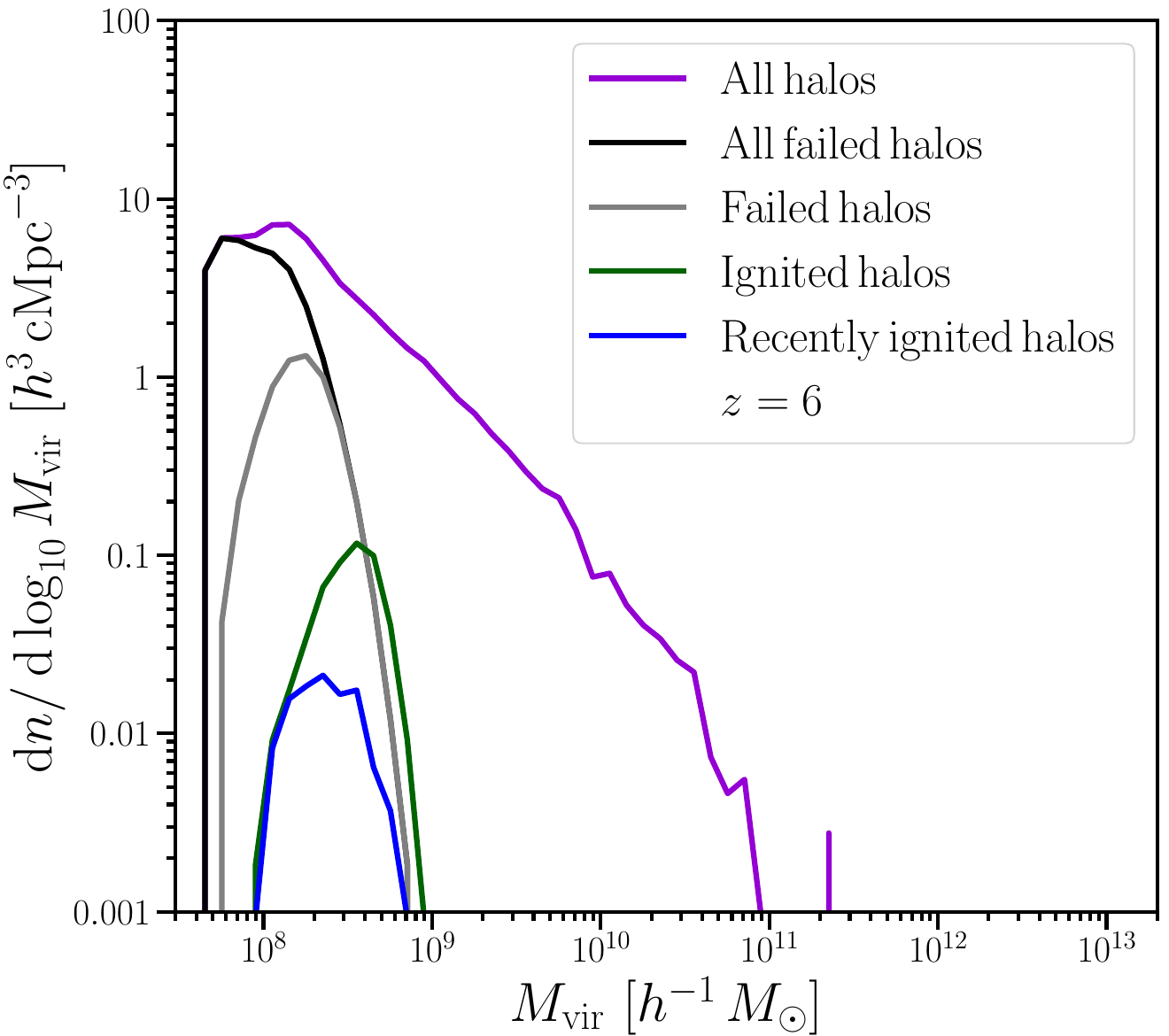}
    \includegraphics[width=0.325\textwidth]{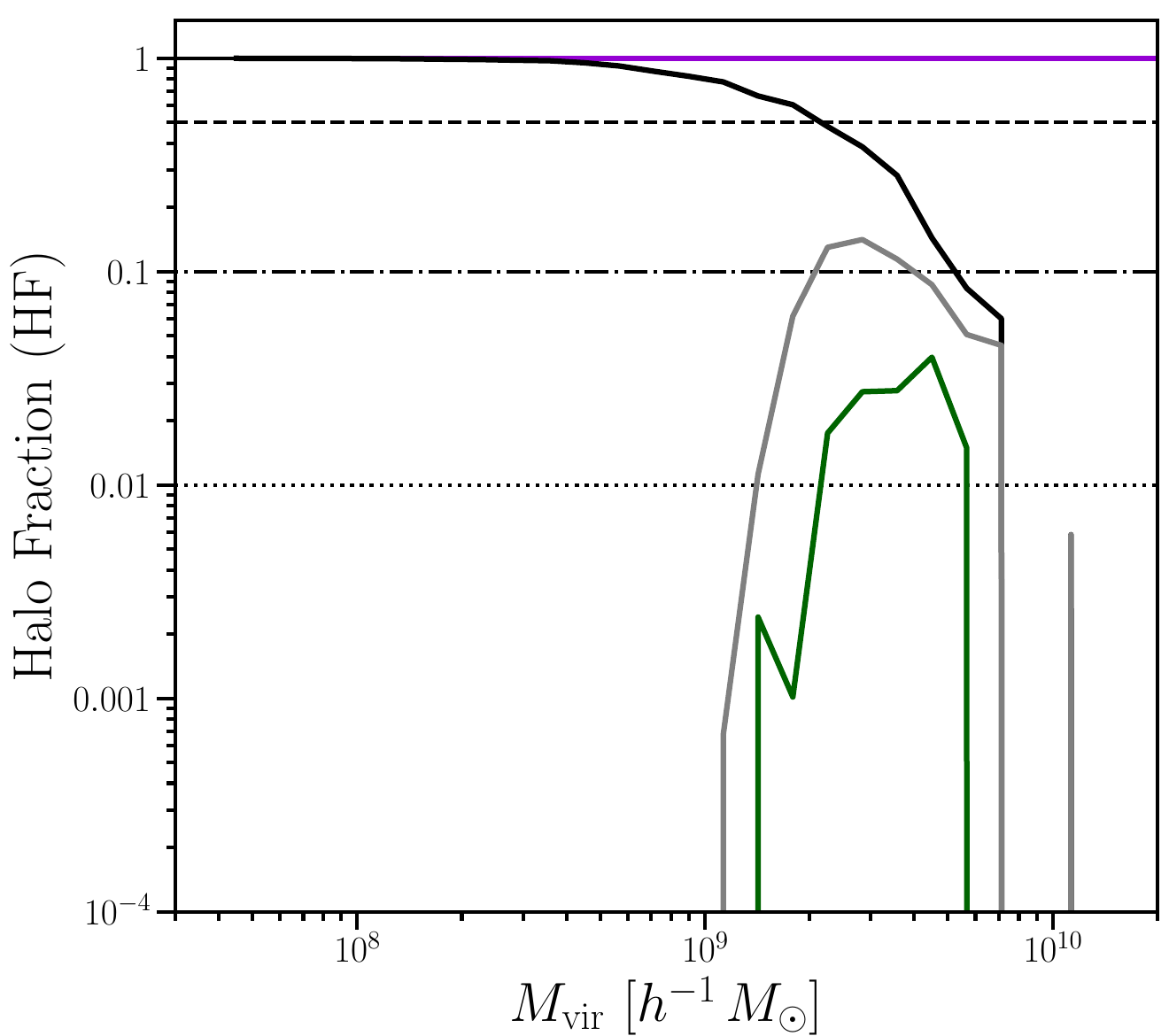}
    \includegraphics[width=0.325\textwidth]{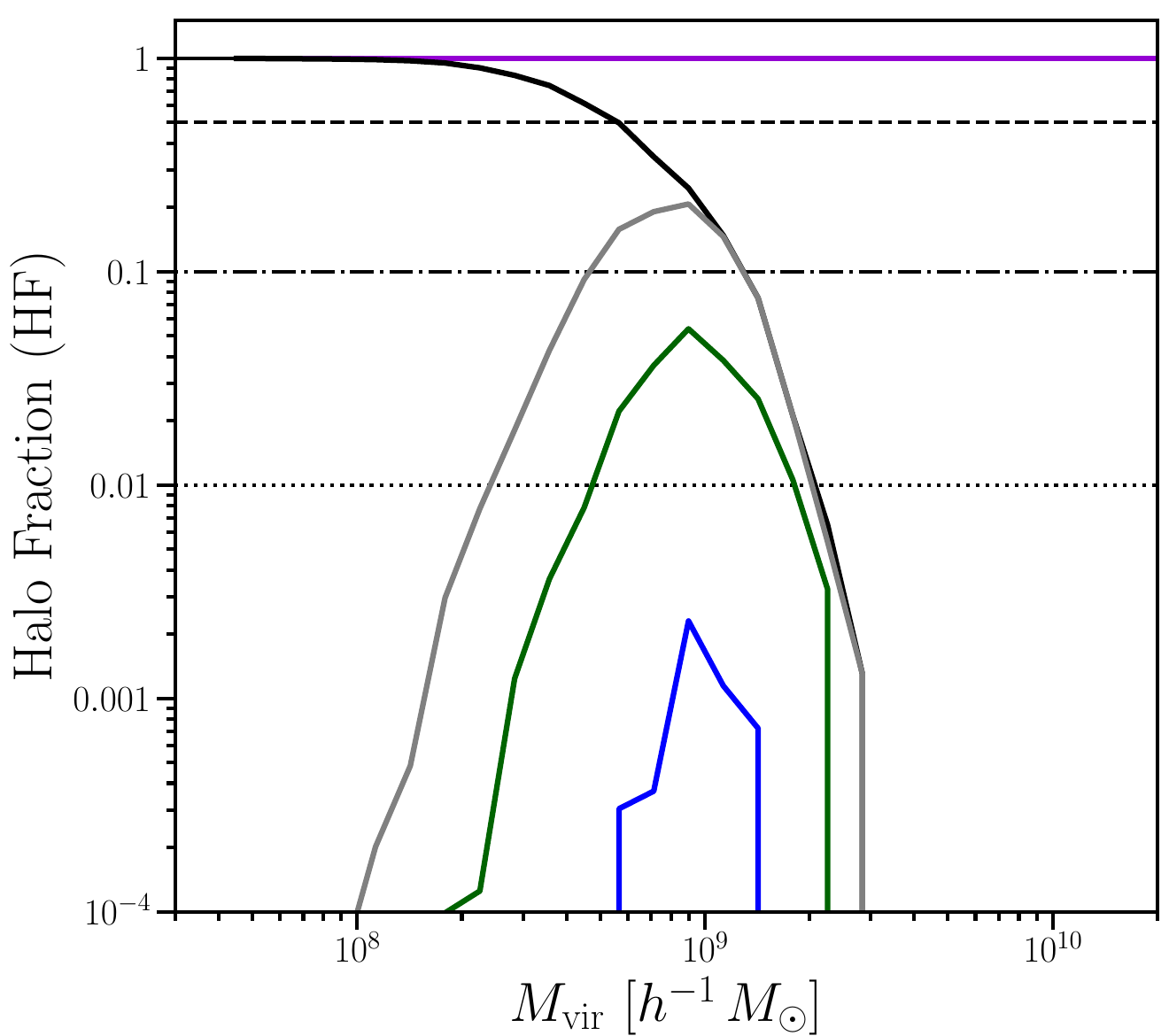}
    \includegraphics[width=0.325\textwidth]{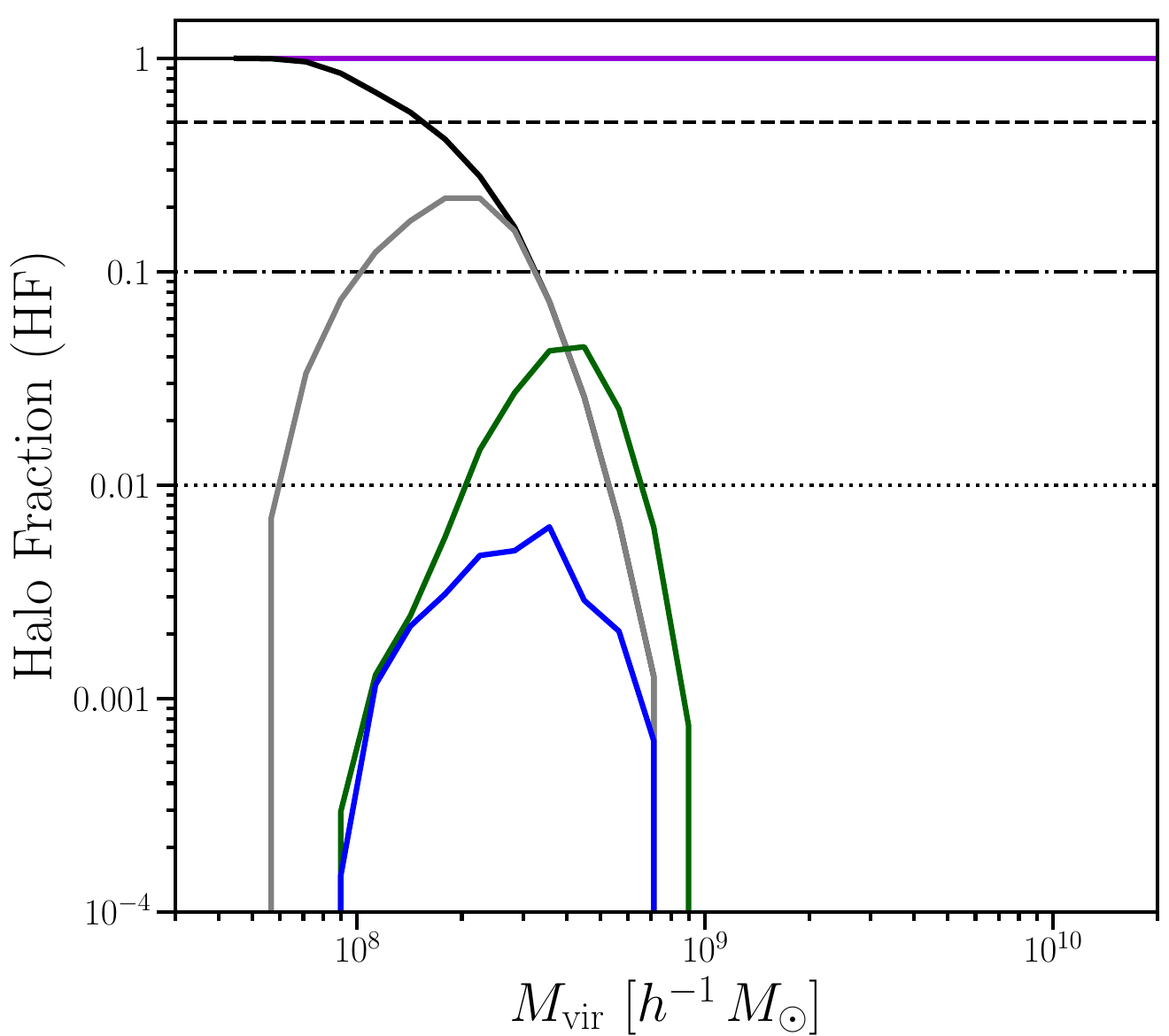}
    \vspace{-0.0cm}    
    \caption{Halo abundances for various populations in our simulation (see Table~\ref{table:samples} for definitions). Violet, black and gray lines represent all, all failed, and failed (gaseous) halos, while green and blue denote ignited and recently-ignited (gaseous) halos. Note: we exclude subhalos from our analysis. {\it Left-to-right panels}: $z = 0, 3$ and $6$. {\it Top panels}: Halo mass functions. {\it Bottom panels}: Halo fractions (HF), defined here as the ratio of the halo mass function of a given population and that of all halos. For the failed, ignited, and recently-ignited populations, the normalization of the halo mass function diminishes with cosmic time. Both halo mass function and HF shift to higher $M_{\rm vir}$ values at lower redshift. This mass-shift aside, the HF levels remain stable with cosmic time for every population -- except for the recently-ignited population, which diminishes quickly with cosmic time, becoming negligible below $z=3$ (i.e., only the blue curve plummets with decreasing redshift). The horizontal dotted, dot-dashed and dashed black lines indicate the $M_{\rm vir}-$thresholds where the various HFs reach $1\%$, $10\%$ and $50\%$ respectively. 
    }
    \label{fig:mass_function}
\end{figure*}

\begin{figure*}
    \centering
    \includegraphics[width=0.497\textwidth]{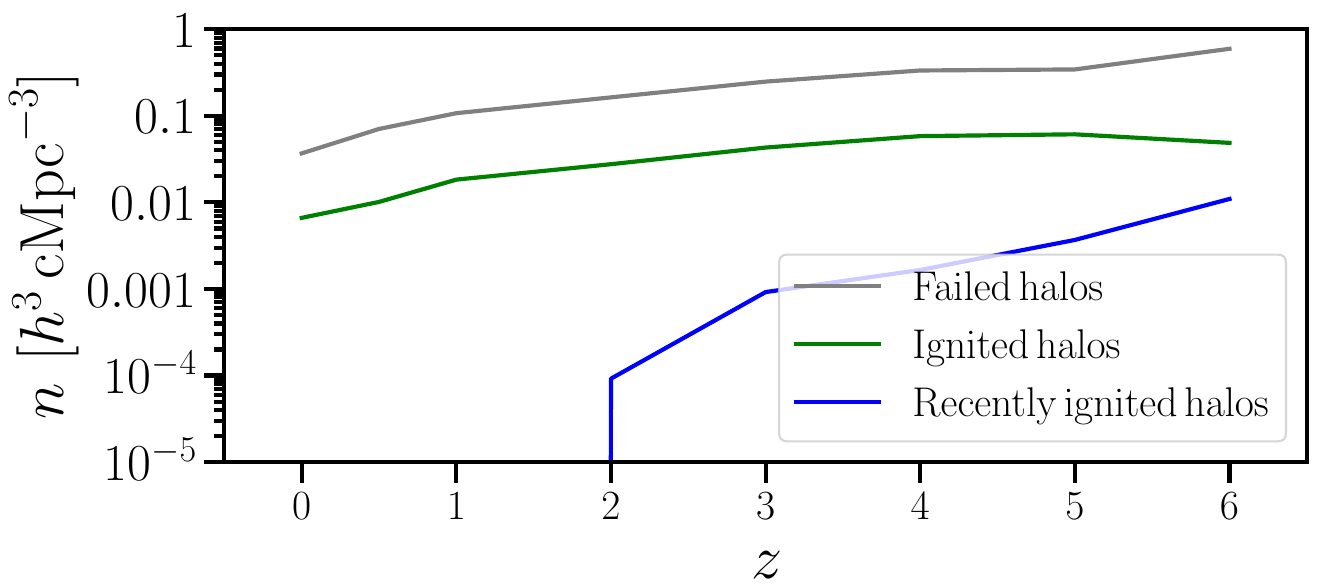}
    \includegraphics[width=0.497\textwidth]{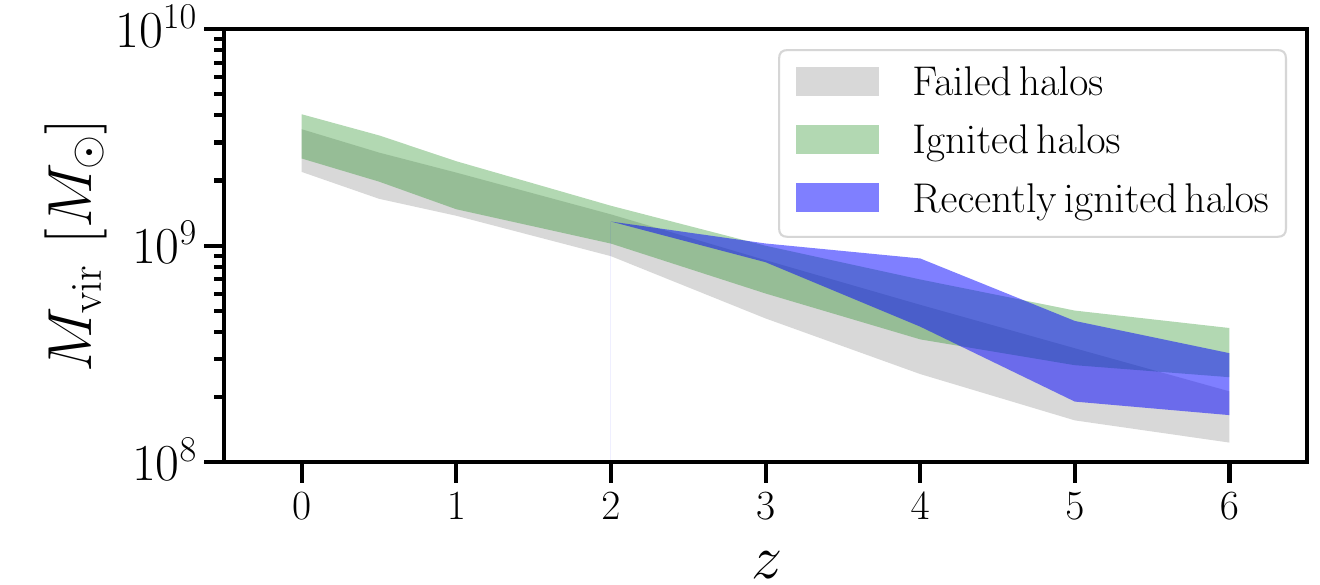}
    \includegraphics[width=0.497\textwidth]{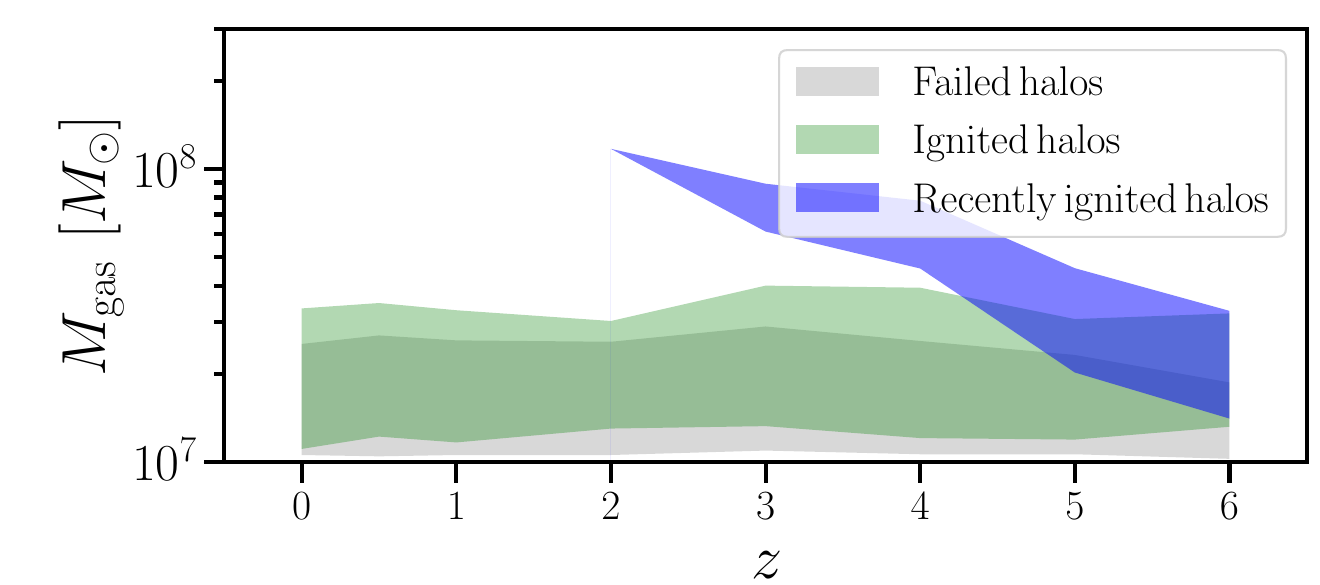}
    \includegraphics[width=0.497\textwidth]{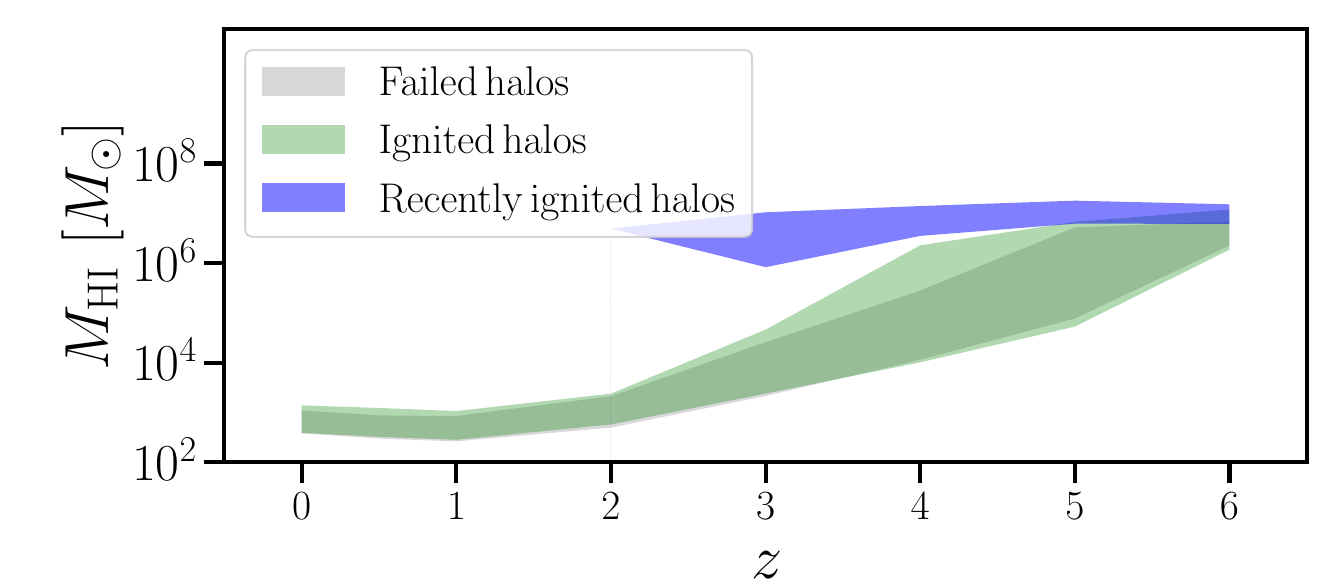}
    \includegraphics[width=0.497\textwidth]{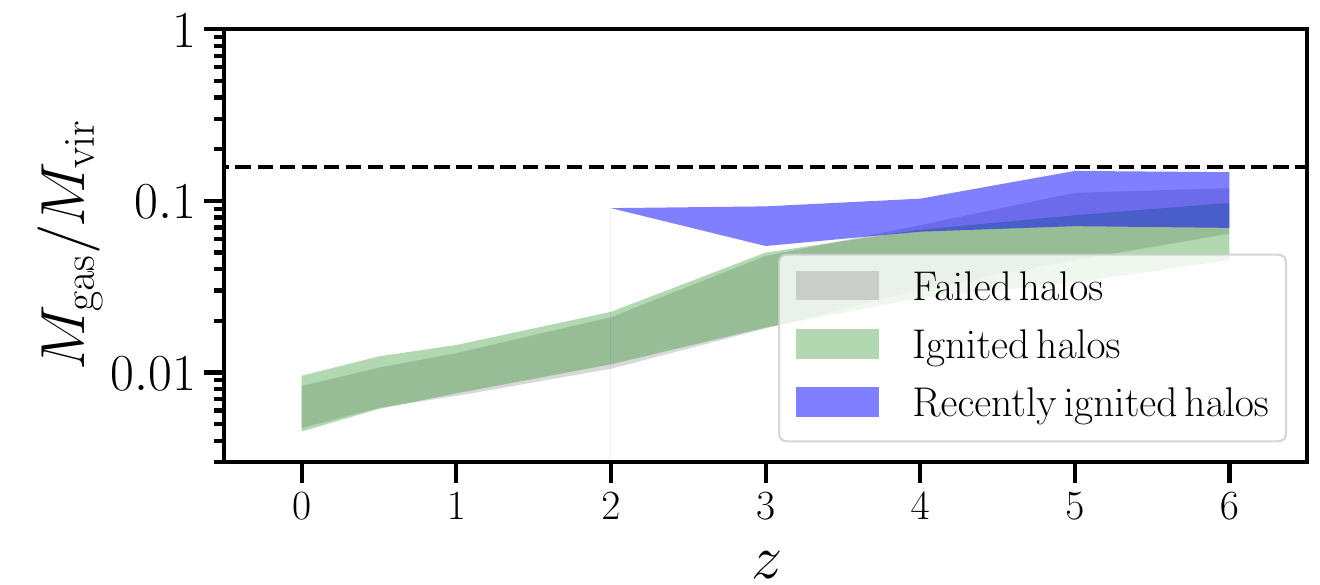}
    \includegraphics[width=0.497\textwidth]{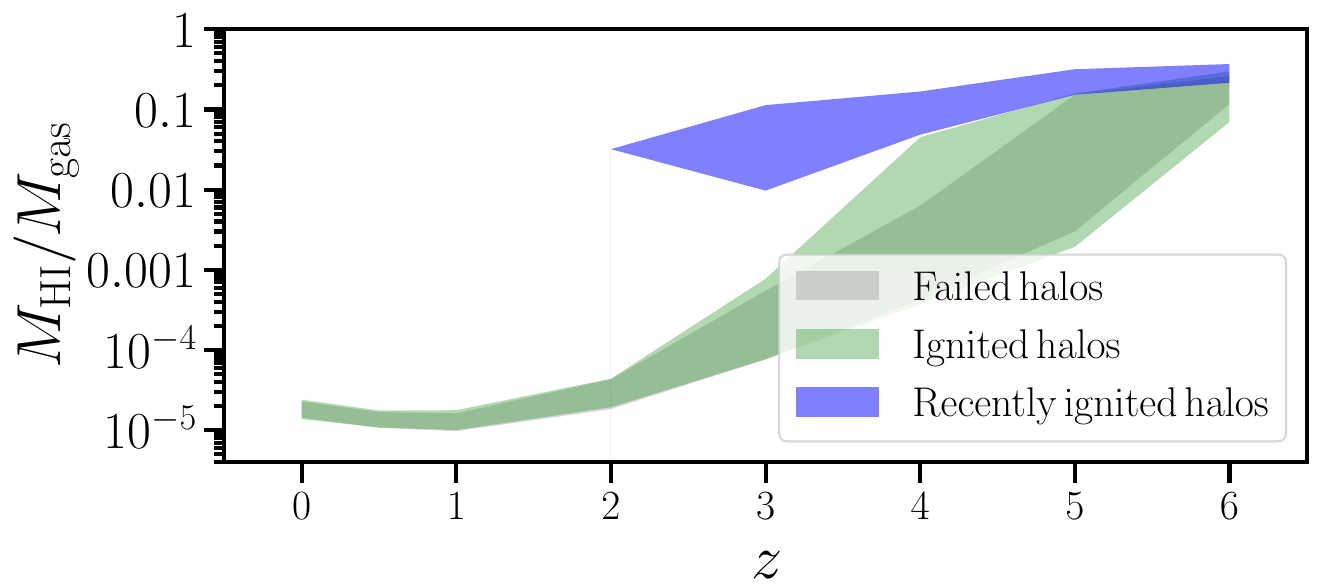}
    \caption{The evolution of the failed, ignited and recently ignited halo populations (gray, green and blue respectively -- see Table~\ref{table:samples} for definitions). Bands represent running medians with widths spanning the first and third quartiles. Recall that we focus exclusively on centrals with a resolved gas component, and our threshold for recent ignition is $t^{\star}_{\rm age}=100$ Myr.  {\it Upper-left to bottom-right:} number density, virial mass, gas mass, HI-gas mass, fraction of mass in gas, and fraction of gas mass in HI. The dashed horizontal black line denotes $\Omega_b/\Omega_m$, the universal gas fraction. Although approximately one order of magnitude less numerous, the ignited population follows their failed counterparts across redshifts (see text for a more thorough discussion of their differences). The recently-ignited population behaves differently. Their abundance plummets faster with cosmic time. Below $z\sim5$, these halos have larger $M_{\rm vir}$ values. However, their gas masses and gas fractions, HI-content and HI-to-gas ratios increase dramatically with cosmic time relative to their older and failed counterparts. Recently-ignited halos tend to have copious gas reservoirs, particularly in the HI phase.
    }
    \label{fig:evolution}
\end{figure*}

\section{Results}\label{sec:results}

\subsection{Halo abundance}\label{subsec:abundance}

Figure~\ref{fig:mass_function} quantifies the abundance of halos of various types considered in this paper: all halos (violet), all failed halos (black), failed halos (gray), ignited halos (green) and recently-ignited halos (blue) for three representative redshifts, $z=0,3$ and $6$ (left-to-right panels) -- see Table~\ref{table:samples} for definitions. Recall that we do not consider subhalos here, and that our focus is on failed or ignited halos with a resolved gaseous component (we drop the `gaseous' adjective for the last three populations mentioned above). The top panels focus on the halo mass function, which, as expected, migrates towards lower $M_{\rm vir}$ values with increasing redshift for every population displayed. We also note that the normalization of the three gaseous populations increases with redshift { (although, for gaseous ignited halos, this stagnates between $z=3$ and $6$)}. This is particularly evident for the recently-ignited halos, which become extremely rare by $z=3$. In our simulation volume, we only find one such object at $z=2$ (described in Figure~\ref{fig:awake_vs_failed}) and none thereafter. 

The bottom panels present the halo fraction (HF), defined as the ratio of the halo mass function of a given population to the `all halos' mass function (i.e., the colored curves `divided' by the violet curves). The dashed, dot-dashed and dotted horizontal lines in these panels indicate where the halo fraction reaches $50\%$, $10\%$ and $1\%$ respectively. We constrain the displayed virial mass range relative to the top panels to emphasize the $M_{\rm vir}-$values covered by the ignited and recently ignited (gaseous) halo samples, plus their failed counterparts (the gray, green and blue curves). These three populations shift towards lower $M_{\rm vir}$ values at higher $z$. For both failed and ignited halos, their overall contribution to the entire halo population remains somewhat stable, with only a mild increase at earlier times. On the other hand, the fraction of recently-ignited halos is always below $1\%$, and diminishes dramatically with cosmic time -- down to a single object in our simulation volume at $z=2$ (see Figure~\ref{fig:awake_vs_failed}), and none thereafter. 

Many authors have recently employed this kind of halo-fraction analysis to pin down a virial mass threshold below which halos no longer form galaxies (see Section~\ref{subsec:threshold} for a detailed discussion). This raises the following question: is $M_{\rm vir}$ the dominant driver in determining whether or galaxy ignition is successful in a halo? Before attempting to answer this question, we inspect the mass evolution of our ignited halos and characterize the nature of their interstellar medium.

\subsection{Evolution and gas content}\label{subsec:gas}

Figure~\ref{fig:evolution} focuses on the evolution of the three halo samples with a resolved gaseous component: the failed, ignited and recently ignited populations (in gray, green and blue respectively -- see Table~\ref{table:samples} for definitions). The upper-left panel shows comoving number densities, and the rest show running medians, with band widths representing the $25^{\rm th}-75^{\rm th}$-percentile range. We emphasize that the trends we discuss below represent how each population manifests at various epochs, not how individual halos evolve across redshifts. Namely, the objects belonging to a given sample at a given time need not belong to that population at later epochs.

The abundance of failed halos decreases by approximately an order of magnitude between $z=6$ and today. The ignited halos are about $1$ dex less abundant at all epochs and follow a similar trend, but with a mild overturn at the highest redshifts. In general, the ignited halos are slightly more massive than their failed counterparts, especially at the earliest epochs. As redshift decreases, the median gas content of both populations remains constant, while their HI budget diminishes. In other words, in general their gas fractions increase with redshift, and their HI-fractions increase more dramatically with $z$. The scatter in $M_{\rm HI}$ and $M_{\rm HI}/M_{\rm gas}$ also increases with redshift. { We remind the reader that these populations only include halos with at least 128 gas particles, rendering the lower-quartile envelops of our results sensitive to resolution.}

The evolution of the recently-ignited population is noticeably different. Their abundance drops more dramatically with cosmic time and disappears completely by $z=2$. At $z=6$, their $M_{\rm vir}$ values split the difference between the other two samples, but become more biased towards larger values at lower redshifts. As cosmic time advances, their median gas content increases dramatically, while their HI budget remains flat. Their median gas fraction and neutral-to-total gas ratio also diminish, but at a milder rate. Indeed, by cosmic noon ($z=2-3$), the few remaining recently-ignited halos tend to contain an order of magnitude more gas and HI than their older and failed counterparts.

In summary, the ignited halo population, although smaller in number, it is not extremely distinct from its failed counterpart (in terms of $M_{\rm vir}$, $M_{\rm gas}$ and $z-$evolution). However, the subsample of recently-ignited objects tends to have higher $M_{\rm gas}/M_{\rm vir}$ and $M_{\rm HI}/M_{\rm gas}$ values. These results suggest that the presence of an HI reservoir in a halo is connected to its ability to form a galaxy. Next we focus on the stellar ages of the recently-ignited halos and explore possible connections with their gas and HI content.


\begin{figure*}
    \centering
    \vspace{-1cm}\includegraphics[width=0.60\textwidth]{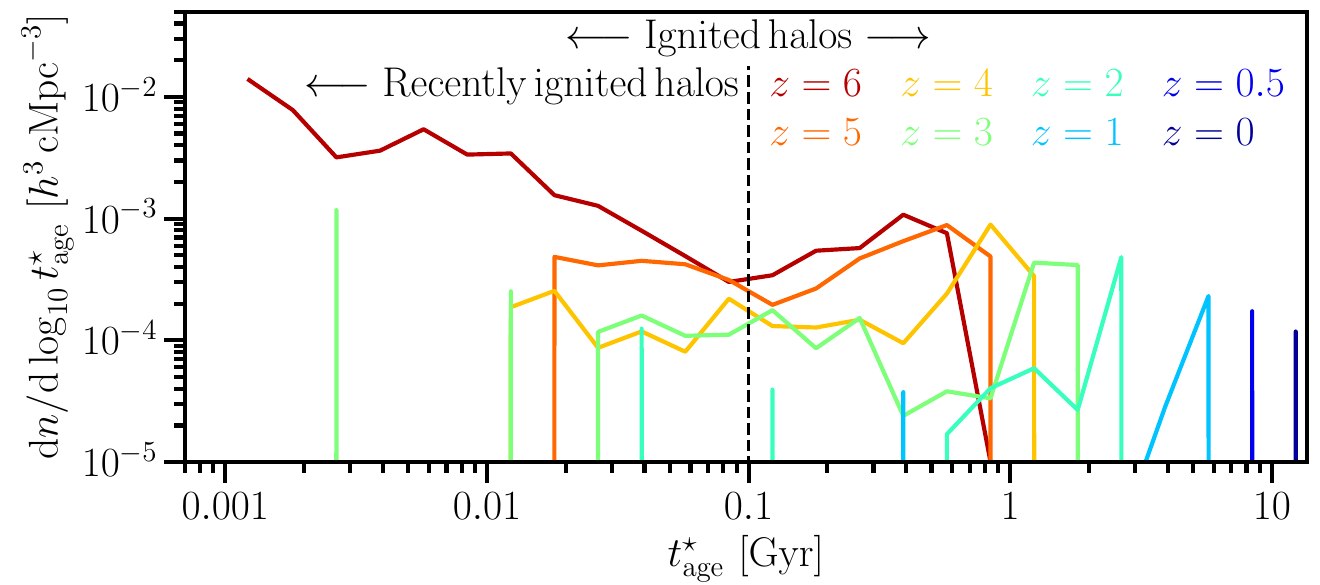}
    \includegraphics[width=0.497\textwidth]{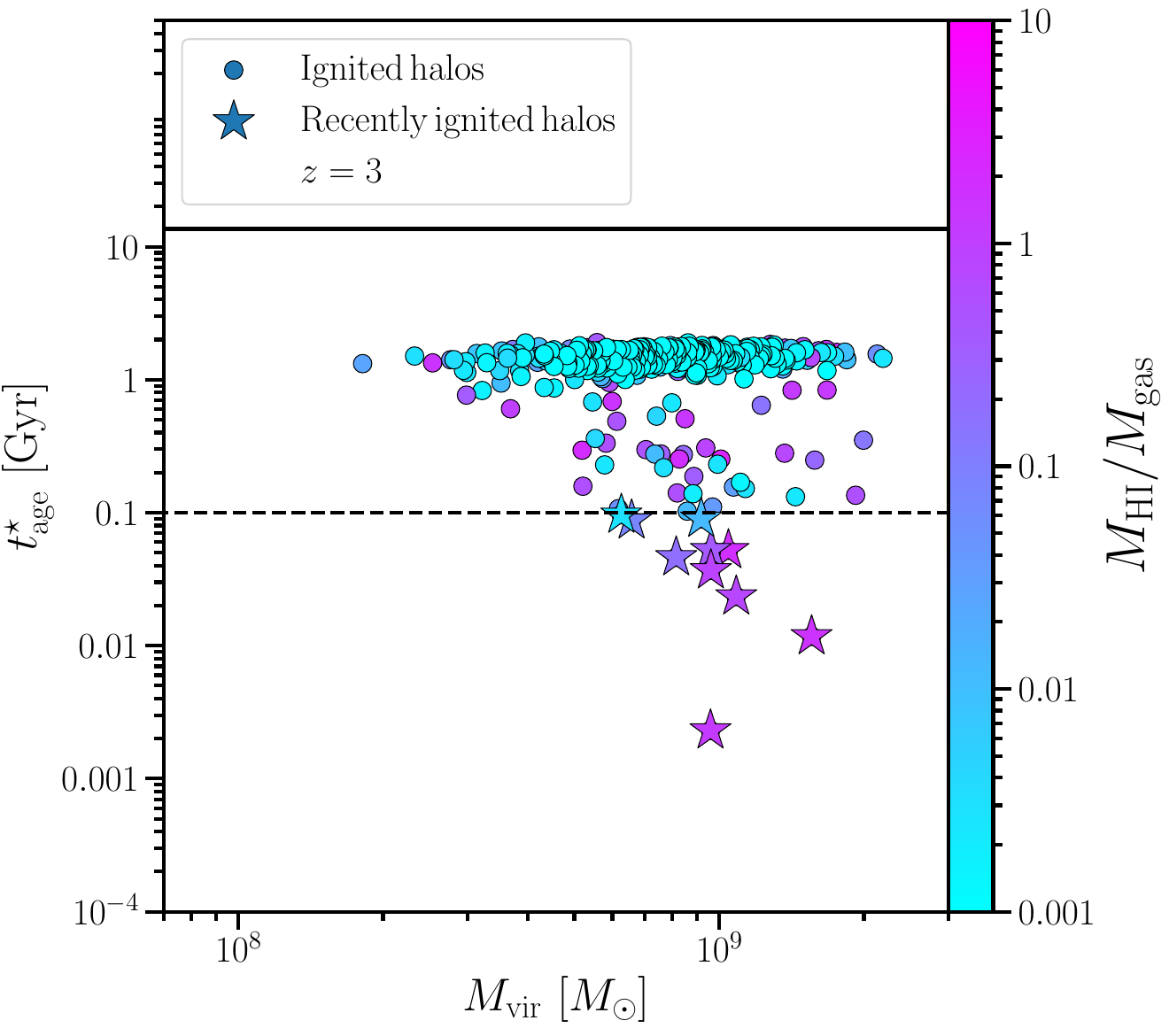}   
    \includegraphics[width=0.497\textwidth]{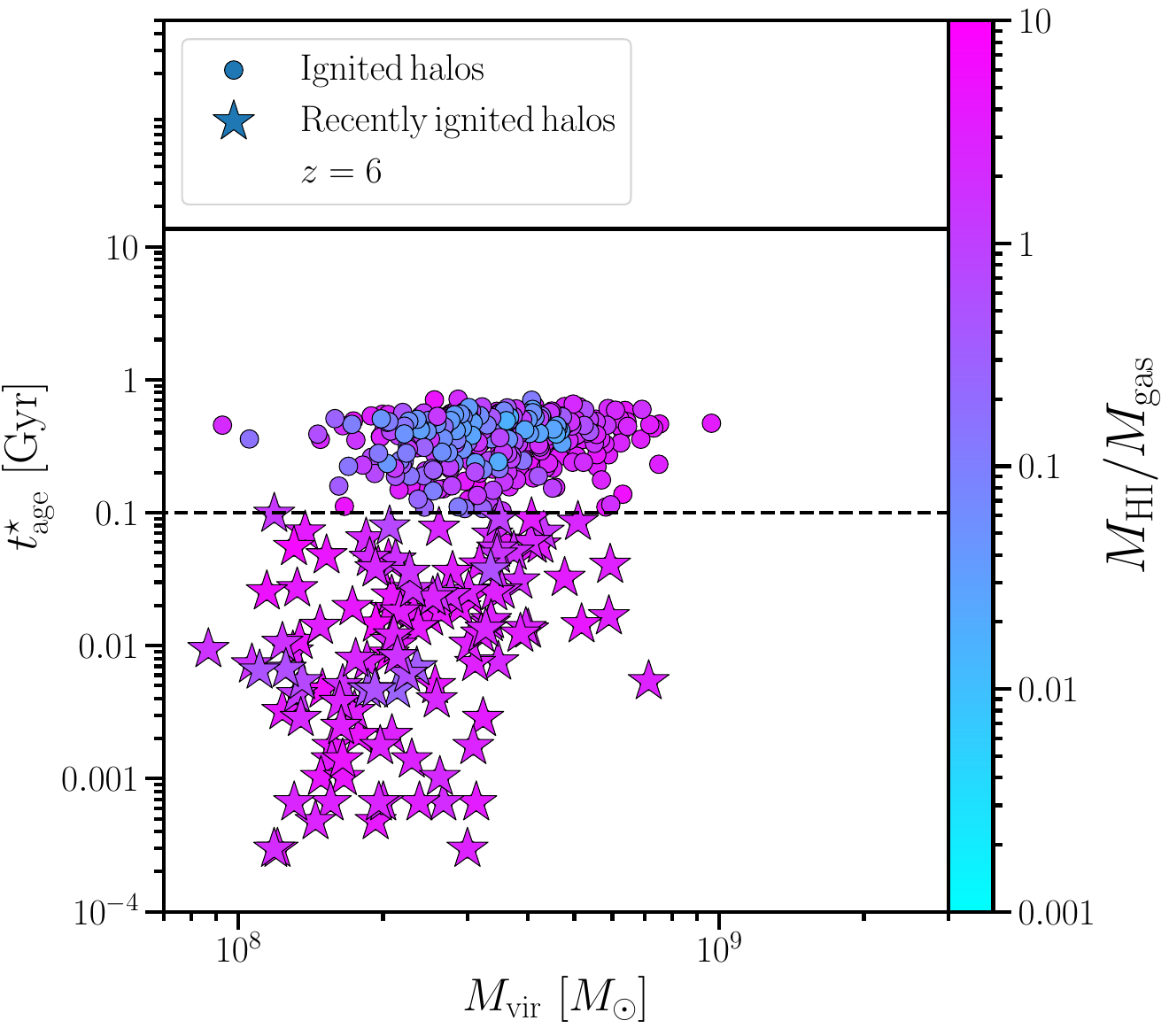}
    \caption{Stellar ages and HI content of ignited and recently-ignited halos (see Table~\ref{table:samples} for definitions). {\it Top panel:} Halo abundance versus stellar ages for the ignited halo population. Each color refers to a different epoch, as indicated by the key. The vertical dashed lane at $t^{\star}_{\rm age} = 100$ Myr demarcates the threshold dividing the recently-ignited halos and their older counterparts. In general, ages increase and abundances decrease with cosmic time. {\it Bottom panels:} Stellar ages of ignited (black circles) and recently-ignited ({ star-shaped symbols}) halos as a function of virial mass, color coded by the HI-to-total gas mass ratio, at $z = 3$ and $6$ ({\it left-to-right}). 
    The horizontal solid and dashed black lines indicate the age of the universe and our $100-$Myr threshold. By $z=3$, the ignited population comes in two varieties: an old and HI-deficient set and a slightly younger sample with a mix of HI fractions, with only a few halos with stellar ages below $100$ Myr. Ignited halos at $z=6$ also break into two populations, though the older sample has stellar ages below $1$ Gyr, with intermediate gas fractions and a mix of intermediate and rich HI content. The younger tail is heavily populated with HI-rich objects. 
    }
    \label{fig:stellar_ages}
\end{figure*}

\subsection{Stellar ages and HI-content}\label{subsec:ages}


The top panel of Figure~\ref{fig:stellar_ages} shows the comoving number density of ignited halos, binned by the age of its stellar population. This is akin to the mass function in the top panels of Figure~\ref{fig:mass_function}, but replacing $M_{\rm vir}$ with $t^{\star}_{\rm age}$. The colors from red to violet represent the eight redshifts considered in this work, as labeled by the upper-right key. The vertical dashed line denotes the $100-$Myr threshold we adopt to select recently-ignited objects from the parent ignited population. In general, the number density normalization diminishes with decreasing $z$ (see also Figures~\ref{fig:cosmic_web} and \ref{fig:mass_function}, plus the upper-left panel of Figure \ref{fig:evolution}). Moreover, populations tend to get older and span narrower (logarithmic) stellar-age ranges. At $z=6$ there is a dominant tail of halos that were ignited between $\sim1$ and $\sim100$ Myr ago, accompanied by another population with ages between $\sim100$ Myr and $\sim1$ Gyr. The populations at $z=5, 4, {\rm and} \, 3$ behave similarly, but with the older population gaining dominance over the younger one. By $z=2$ there is only one recently-ignited halo left in the entire simulation volume (see also Figure~\ref{fig:awake_vs_failed}). 

The subsequent bottom panels also display $t^{\star}_{\rm age}$, but now in terms of \,$M_{\rm vir}$, and color-coded by $M_{\rm HI}/M_{\rm gas}$. We focus only on $z=3 \, {\rm and} \,\, 6$, spanning the range of redshifts we analyze here that exhibit a significant population of recently-ignited halos (these two epochs bracket the behaviour present in the other redshifts we inspect). The solid and dashed horizontal black lines denote the age of the Universe and our $100$ Myr threshold separating the recently-ignited halos ({ star-shaped symbols}) from the general ignited population (black circles). The $z=3$ case shows a prominent HI-poor population with stellar ages of $\sim 1-2$ Gyr, plus a tail of younger halos with a mix of HI-to-total gas ratios. The recently ignited sample is HI-rich, and is slightly biased towards a more massive and narrower range of $M_{\rm vir}-$values. At $z=6$, on the other hand, the older subpopulation becomes younger and shifts from low to intermediate HI content, while the younger population remains HI-rich and extends to lower $t^{\star}_{\rm age}$ values (see also the lower-right panel of Figure~\ref{fig:evolution}). The scatter in $M_{\rm vir}$ for both populations at this redshift is comparable, though the recently-ignited subsample tend to have lower virial masses (see also the upper-right panel of Figure~\ref{fig:evolution}).

At every redshift we inspect (including those not shown in Figure~\ref{fig:stellar_ages}),  halos with $t^{\star}_{\rm age}>100$ Myr tend to have a mix of low and intermediate HI masses, while their younger recently-ignited counterparts tend to be HI-rich. Overall, we find no evidence of a correlation between either stellar age or HI-content with virial mass. Similarly, we also do not find a connection between these two quantities and total gas mass (not shown). Therefore, this motivates us to investigate galaxy ignition beyond $M_{\rm vir}$ and $M_{\rm gas}$ thresholds and to compare the ISM properties of recently ignited halos with their failed counterparts. However, care must be taken because these two populations occupy slightly different regions on the $M_{\rm vir}-M_{\rm gas}$ plane (Figure~\ref{fig:evolution}), and said regions also evolve with $z$. Thus, for a meaningful comparison, for each recently-ignited halo it is essential to compare it to a control set of carefully-matched failed halos.

\begin{figure*}
    \centering
    \includegraphics[width=0.497\textwidth]{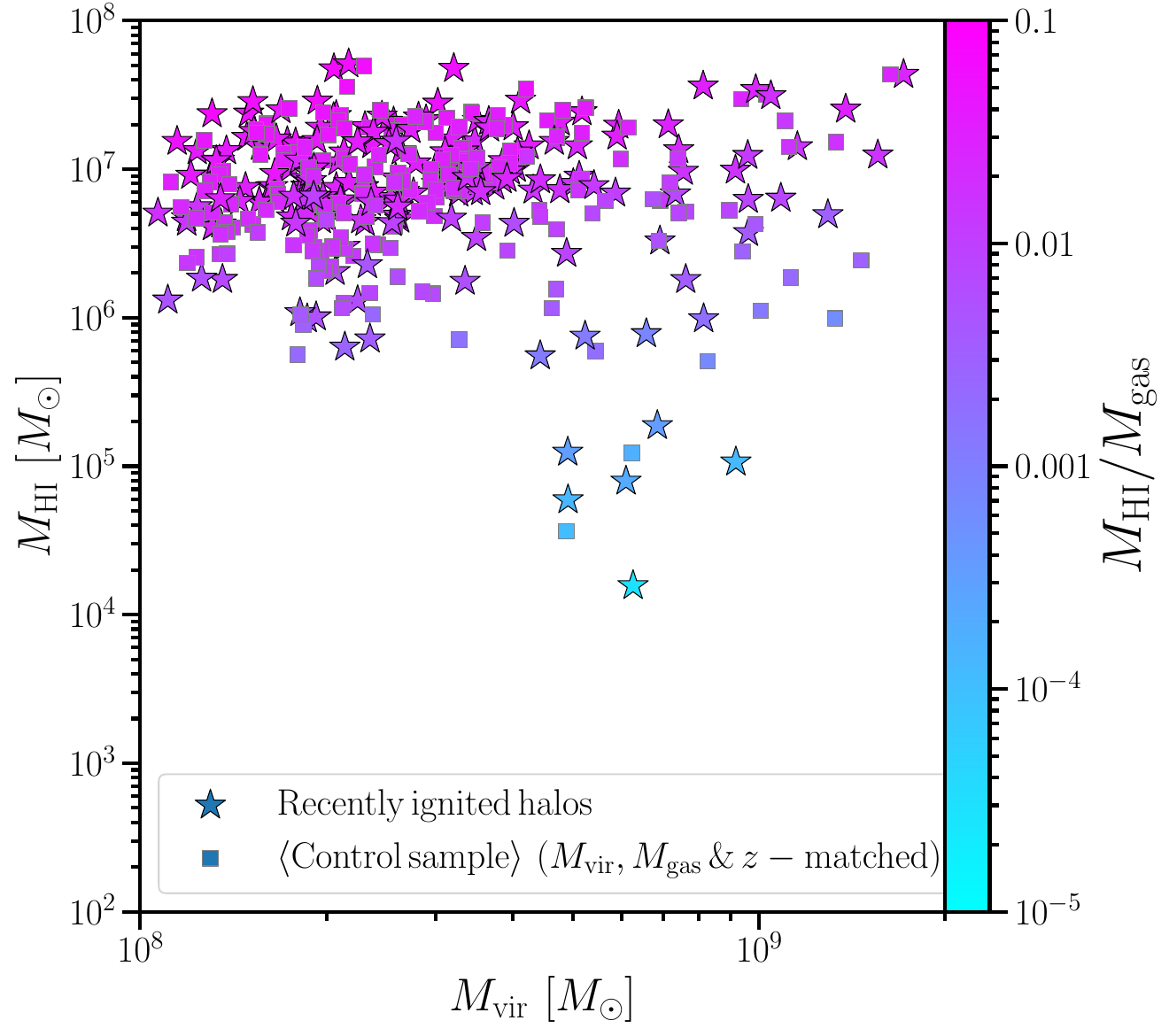}
    \includegraphics[width=0.497\textwidth]{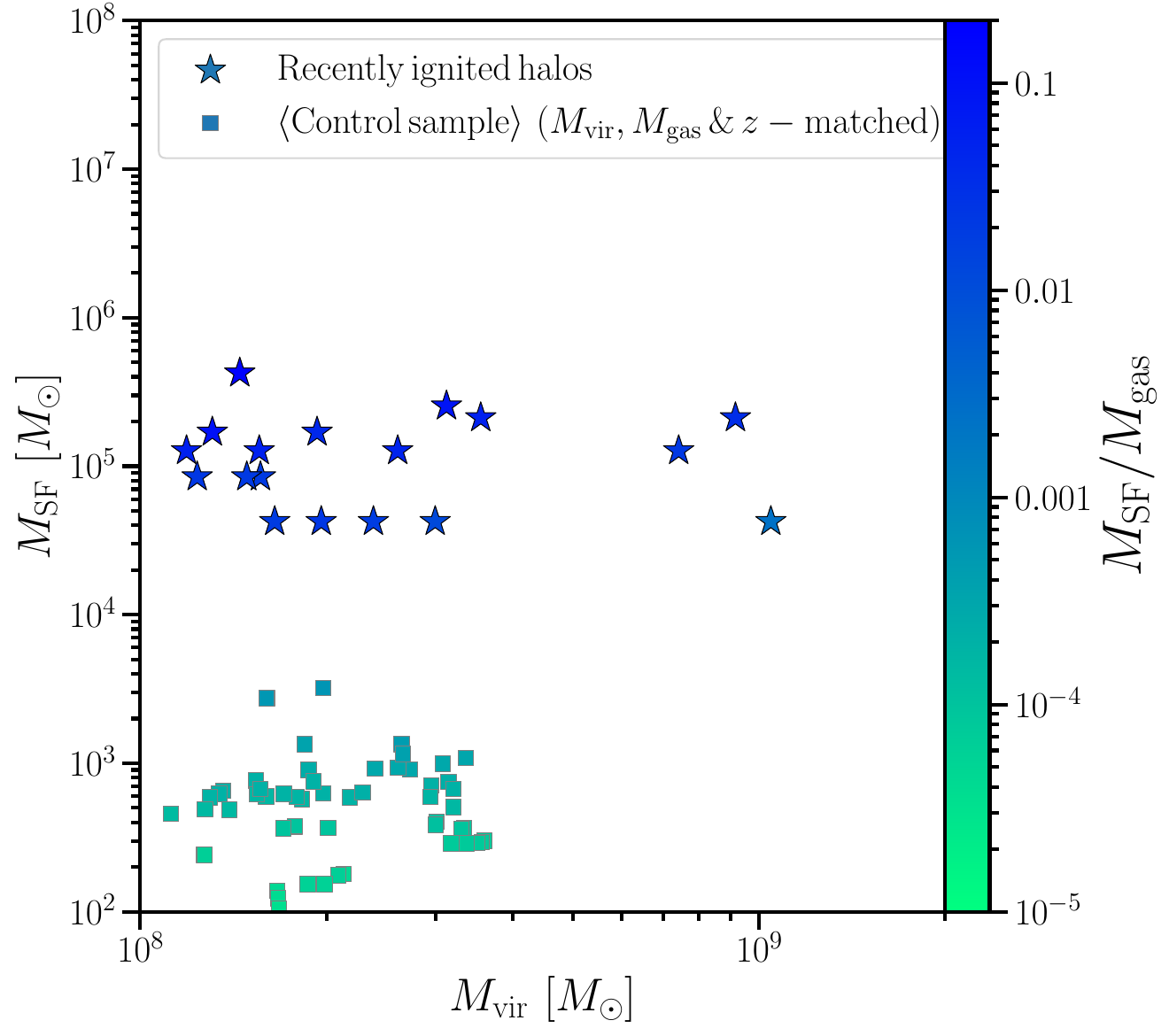}
    \includegraphics[width=0.497\textwidth]{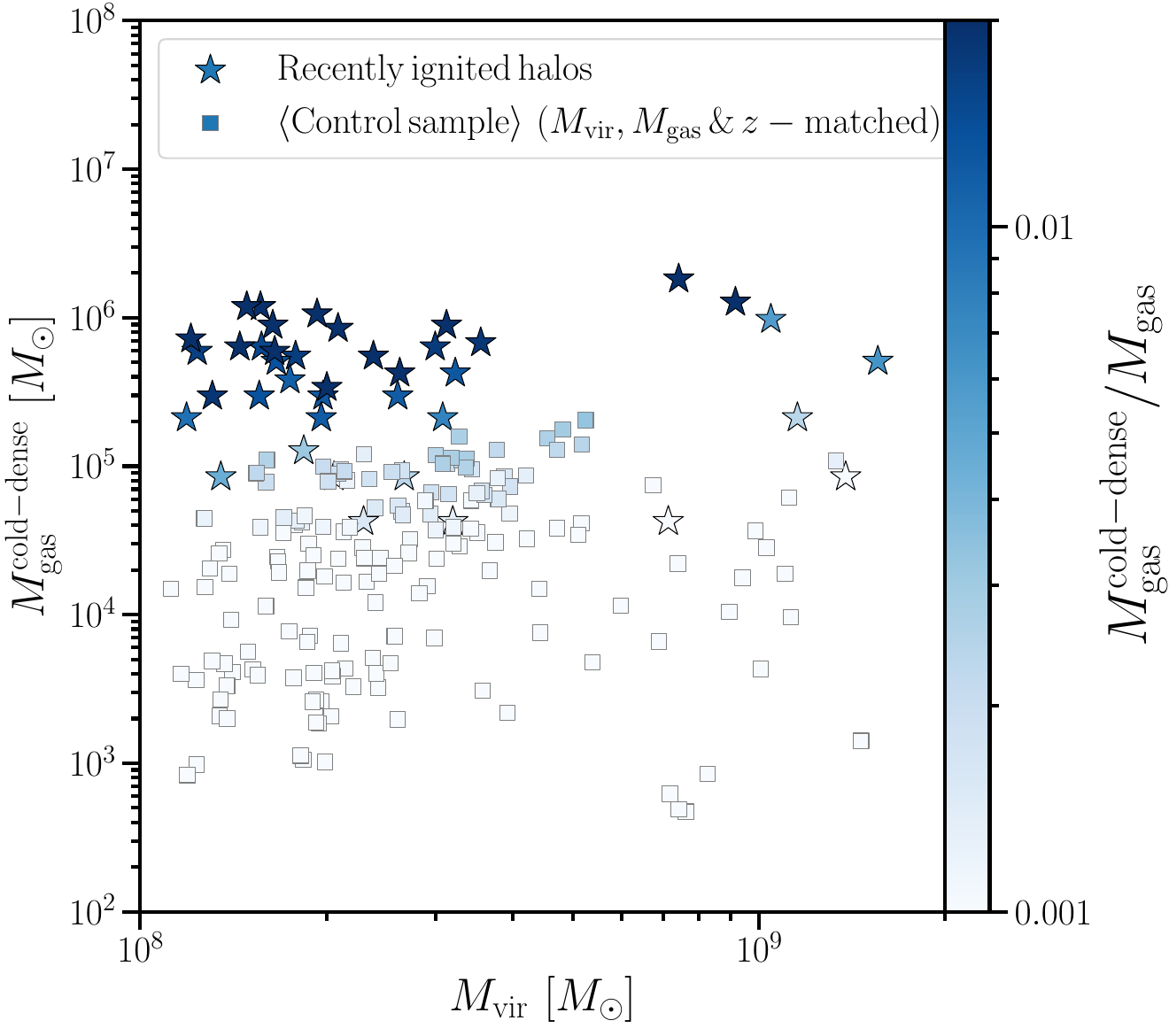}
    \includegraphics[width=0.497\textwidth]{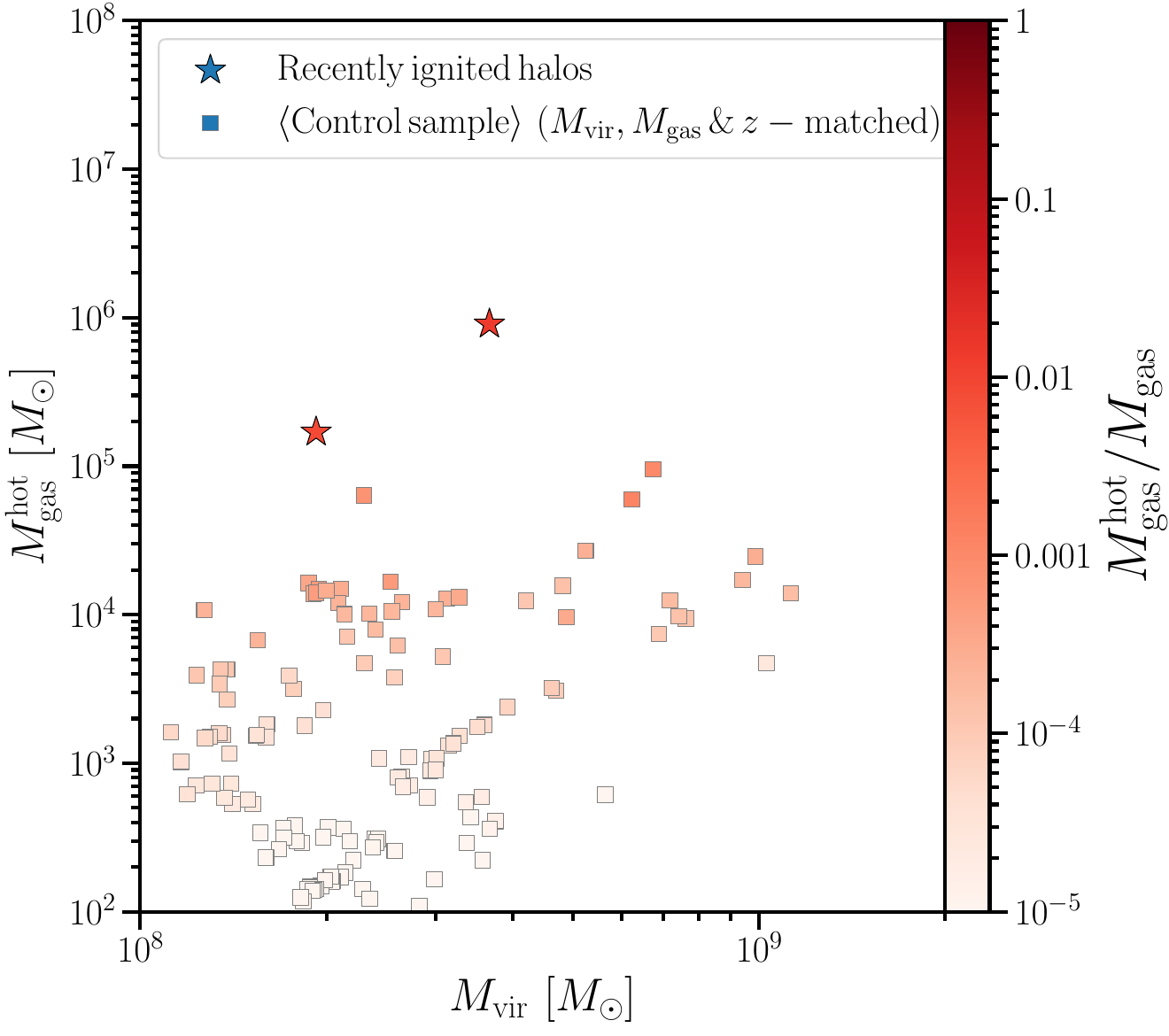}
    \caption{The interstellar medium of recently ignited halos. {\it Top-left-to-bottom-right:} HI, star-forming gas, { cold-dense gas (temperature $< 10^2$ K and density $> 10^2$ cm$^{-3}$) and hot gas (temperature $> 10^6$ K).} Each panel shows the mass in a given ISM versus virial mass, color coded by the fraction of gas mass in that specific phase. The { star-shaped symbols} denote our recently-ignited halos and the { squares} represent average value for each set of failed control halos, matched by virial mass, total gas mass and redshift. The samples we display are the compilation of detections in our simulation at the following redshifts: $z=2, 3, 4, 5$ and $6$ (we exclude $z=0, 0.1$ and $1$ because no recently-ignited halos are identified at those epochs). The HI content in both populations is largely indistinguishable from each other. A handful of recently-ignited halos have star-forming gas, with 2$-$3 orders of magnitude in mass in this component relative to their control failed counterparts. The recently-ignited halos tend to be, for the most part, rich in cold-dense gas, unlike their failed control counterparts. Lastly, across our sampled redshifts, we only find two recently ignited halos with hot gas -- whereas hot-gas content is common among their matched failed halos.
    }
    \label{fig:control_gas}
\end{figure*}

\subsection{The interstellar medium}\label{subsec:properties}

The interstellar medium of a galaxy is an essential ingredient in governing its star formation \citep{Kennicutt1998}. Indeed, any process that augments the colder and denser portions of ISM has the potential to enhance that galaxy's star formation levels \citep{Moreno2019,Moreno2021,He2023}. This elicits the following pertinent question: what conditions must the ISM\footnote{Herein we casually use the term ISM to describe the (potentially) multi-phase gas medium of halos devoid of stars or with a single stellar particle -- even though the ISM is, strictly speaking, the medium {\it in-between} the stars.} of a halo devoid of stars fulfill to successfully ignite galaxy formation? 

To address this question, we compare the ISM of recently-ignited (gaseous) halos against that within failed (gaseous) halos (see Table~\ref{table:samples} for definitions). However, recall that these two populations occupy slightly distinct regions on the $M_{\rm vir}-M_{\rm gas}$ plane, and their differences evolve with redshift (Figure~\ref{fig:evolution}). This means that comparisons in bulk between the two populations might yield misleading insight \citep[see e.g.,][]{Lee2024,Jeon2025}. Concretely, consider some quantity $x$ that appears to be associated with galaxy ignition. If this quantity is strongly connected to either $M_{\rm vir}$ and/or $M_{\rm gas}$, it is then difficult to decide whether $x$ itself is driving galaxy-ignition or $x$ happens to be in a particular $M_{\rm vir}$ (or $M_{\rm gas}$) regime where galaxy-ignition is particularly favorable.

To circumvent this, we compare recently-ignited halos to a control\footnote{This control-based approach is also commonplace in investigations on the role of galaxy interactions in enhancing star formation \citep[e.g.,][]{Ellison2008,Moreno2015,Patton2016,Moreno2019,Moreno2021}.} set of carefully-matched failed halos. In practice, for each recently-ignited object, we search for failed halos at that same redshift whose $M_{\rm vir}$ and $M_{\rm gas}$ are within $0.1$ dex of the target halo. This specific choice is large enough to guarantee that each target (recently-ignited) halo has at least one failed control counterpart, but still smaller than the scatter in $M_{\rm vir}$ and $M_{\rm gas}$ for either population -- see Figure~\ref{fig:evolution}. 

This target-versus-control comparison is appropriate for a second reason. Consider a quantity $y$ for which both recently-ignited and failed halo populations overlap entirely. One might naïvely conclude that this $y$ property plays no role in galaxy ignition \citep[e.g.,][]{Lee2024,Jeon2025}. However, if instead we compare recently-ignited target halos against carefully-matched failed controls, a difference in $y$ values might emerge on an object-by-object basis. Below we discuss one concrete example of this situation.

Figure~\ref{fig:control_gas} illustrates this exercise. In all panels, the star-shaped symbols represent the recently-ignited population and the squares denote the result of averaging over their matched failed counterparts. We incorporate every redshift containing recently-ignited halos ($z=2, 3, 4, 5$ and $6$). This is appropriate because we also match by redshift in our target-control comparison. The top-left panel shows HI mass versus virial mass, color coded by the fraction of gas in HI. We find that every recently-ignited halo contains some HI gas, and each of these halos also has at least one matched failed object also containing HI -- although some of the reported averages involve calculations that may include matched halos without any HI gas. We do not identify any significant segregation between the two populations. Namely, the prevalence of HI gas in recently-ignited objects relative to failed halos (Figure~\ref{fig:evolution}) vanishes on a population-wide level once we match by $M_{\rm vir}$, $M_{\rm gas}$ and $z$. Nevertheless, this need not be true on an object-by-object basis (more below).

Note that we are reporting the average value for each control set corresponding to a given recently-ignited halo. For example, the top panels of Figure~\ref{fig:awake_vs_failed} shows the only recently-ignited halo at $z=2$ in our simulation volume, with $M_{\rm HI} \simeq 4.94 \times 10^6 M_{\odot}$. However, the bottom panels of Figure~\ref{fig:awake_vs_failed} show only one of the nine control failed halos matched to this object. This failed halo has $M_{\rm HI} \simeq 5.75 \times 10^4 M_{\odot}$, while the median value among this set of nine is $\simeq 1.39 \times 10^5M_{\odot}$ (with $25-75$th percentiles of $\simeq 0.42 -7.85 \times 10^5M_{\odot}$). I.e., for this particular case, the median HI content of the failed control set is $\sim1.5$ dex lower than its recently-ignited counterpart, and has a scatter of $\sim1.3$ dex, comparable to what is reported in the upper-left panel of Figure~\ref{fig:control_gas}.

\begin{figure*}
    \centering
    \includegraphics[width=0.49\textwidth]{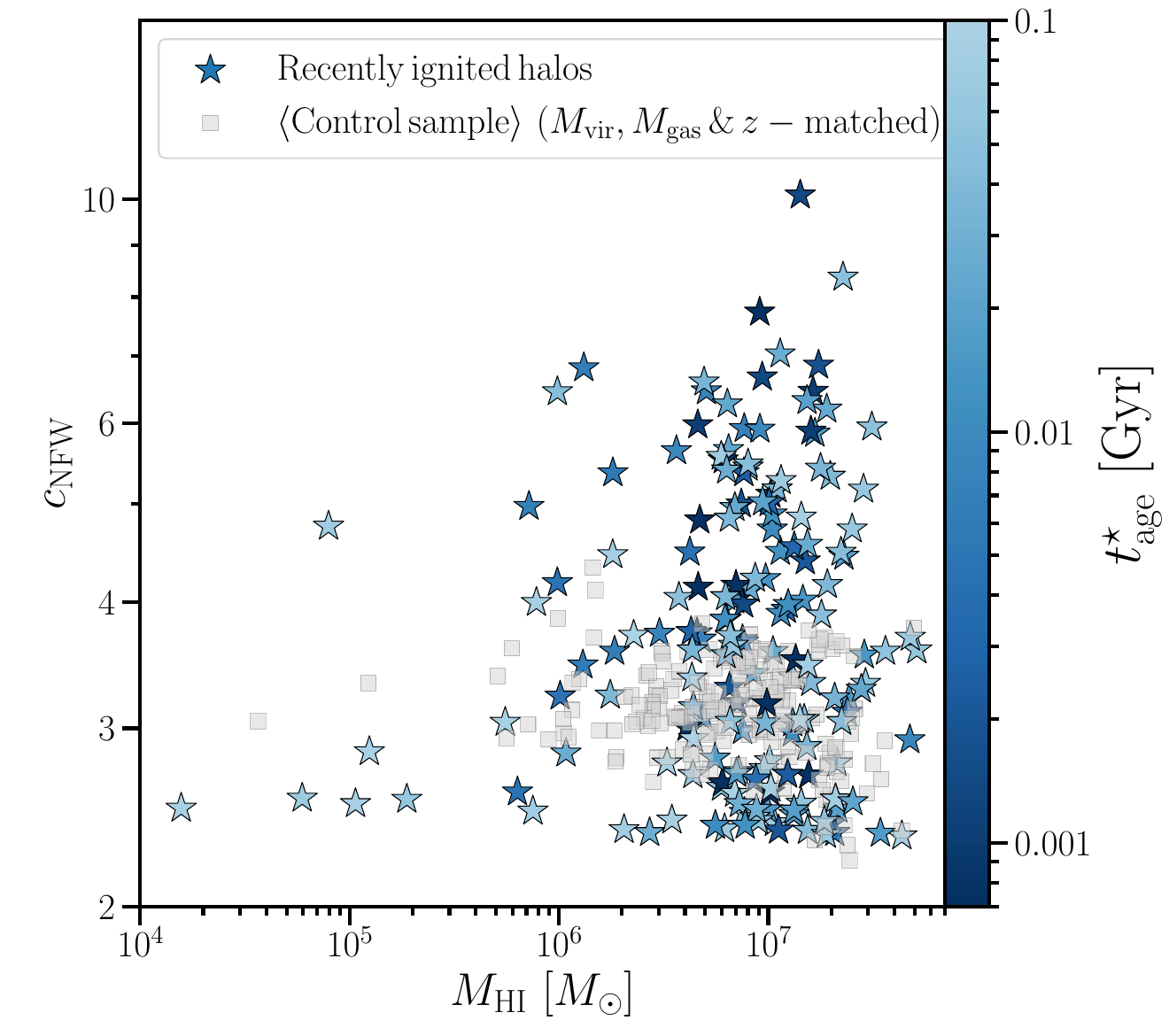}
    \includegraphics[width=0.49\textwidth]{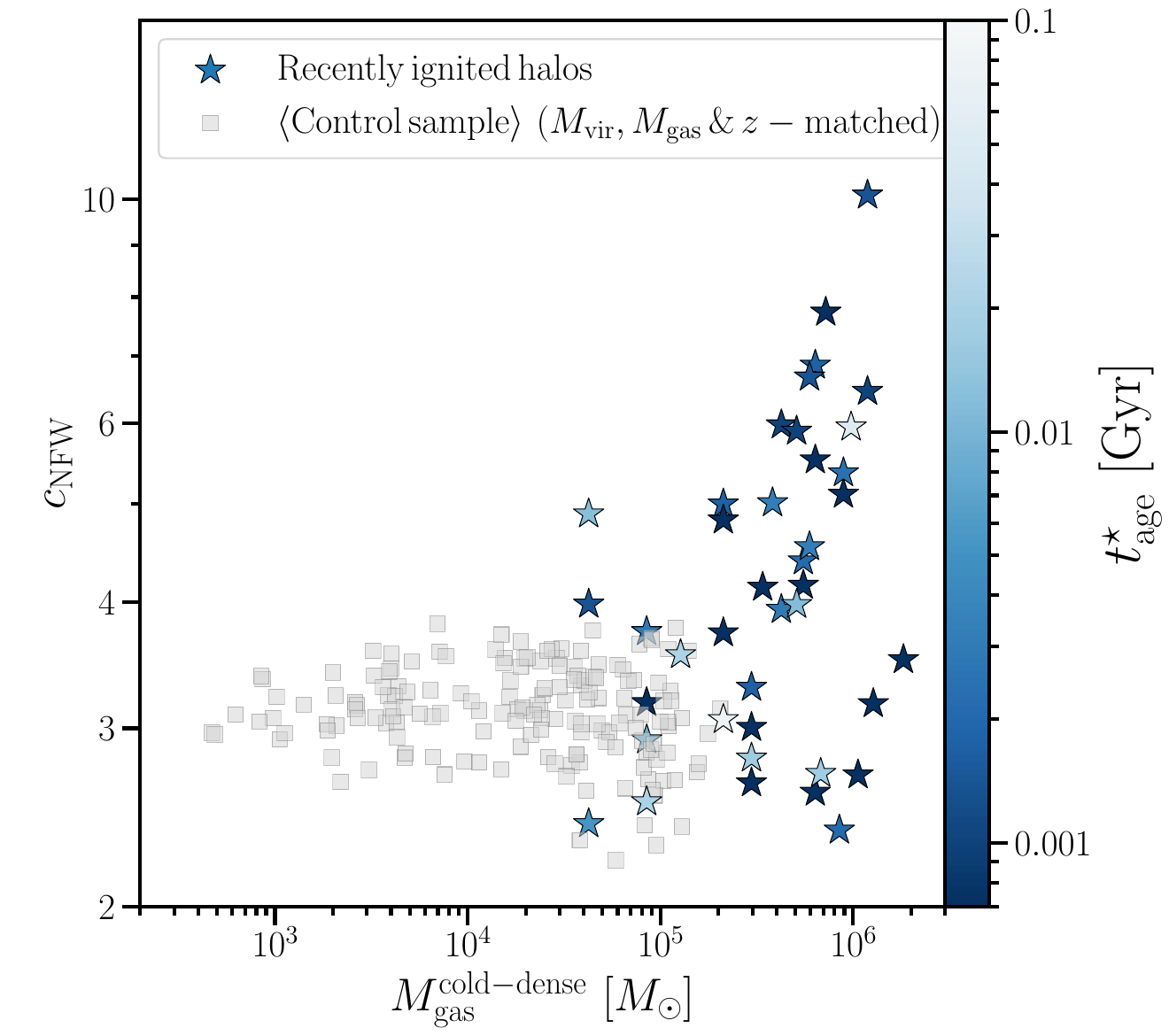}
    \caption{ISM and halo properties as potential drivers of galaxy ignition. NFW concentration versus HI gas mass ({\it left panel}) and cold-dense gas mass ({\it right panel}) for recently-ignited halos ({ star-shaped symbols}, color-coded by stellar age) and failed halos matched by $M_{\rm vir}$, $M_{\rm gas}$ and $z$ (gray squares denoting average values for each control set). The range of $M_{\rm HI}$ values overlaps for both populations, but the recently-ignited sample contains a subpopulation of halos with much higher concentrations ($c_{\rm NFW} \sim 4-10$). In terms of cold-dense gas content, the failed control sample tends to have lower gas budgets, allowing for a clearer segregation between these two populations.
    }
    \label{fig:cnfw}
\end{figure*}

The upper-right panel of Figure~\ref{fig:control_gas} shows this comparison in terms of gas eligible for star formation according to our \texttt{FIRE-2} physics model (hereafter `star-forming' gas). The difference between these two populations is striking, with the recently-ignited halos containing $\sim2$ orders of magnitude more star-forming gas than their matched failed counterparts. Note that the number of symbols here is lower than in the previous panel. This is caused by two reasons: not every recently-ignited halo has star-forming gas, and not every recently-ignited halo (regardless of star-forming content) has a control set containing at least one failed halo with star-forming gas. Concretely, in this simulation volume, only $\sim9\%$ of the recently-ignited halos contains star-forming gas. Likewise, only $\sim28\%$ of the matched-control collections contain at least one failed halo with star-forming gas. 

Another way to characterize the ISM is by slicing the gas temperature-density phase diagram into regions. Following \cite{Moreno2019}, we focus on two contrasting regions: cold-dense gas (temperature $< 10^2$ K and density $> 10^2$ cm$^{-3}$) and hot gas (temperature $> 10^6$ K) -- see \cite{Martizzi2019} for an alternative slicing approach. The bottom panels of Figure~\ref{fig:control_gas} compares these two ISM phases between our recently-ignited and matched-failed populations. We find that the cold-dense gas content of recently-ignited halos tends to exceed that in their matched-failed counterparts by $\sim1-2$ dex. However, this statement only applies to those objects with some cold-dense gas, amounting to only $\sim21\%$ of the entire population. Likewise, among the sets of matched-controls, only $\sim88\%$ have a failed halo with some cold-dense gas. 

It is very rare for a recently-ignited halo to contain hot gas ($1\%$) -- but those few objects with a hot gas component have substantially more mass in this phase than their matched-failed counterparts (dark red-filled star-shaped symbols versus squares with lighter red hues). It is possible that supernovae explosions are the culprit behind such high gas temperatures. We plan to explore this possibility in a future paper. On the other hand, a large fraction of matched-control collections have at least one failed halo with hot gas ($\sim70\%$). Although the presence of hot gas is not necessarily a good tracer of recent ignition, our results suggest halos detected to have $\sim10^5-10^6M_\odot$ of hot gas are more likely to have formed an ultra-faint galaxy recently. 

Overall, after controlling by $M_{\rm vir}$, $M_{\rm gas}$ and $z$, the recently-ignited and matched-failed halo populations are strongly segregated from one another in terms ISM properties (except when HI mass is considered). However, this approach has two limitations: not every recently-ignited halo has a given ISM phase and not every set of matched-controls has at least one failed halo with said ISM phase. Therefore, although the properties of the ISM might be a powerful signposts for recent galaxy ignition, such imprints might not always be available. 

Beyond the properties of the ISM, we may pose another question: what other characteristics help drive the ignition of galaxy formation in otherwise starless halos (with a resolved gas component)? To address this, we next focus on the halo structure of these objects.

\begin{figure*}
    \centering
    \includegraphics[width=0.49\textwidth]{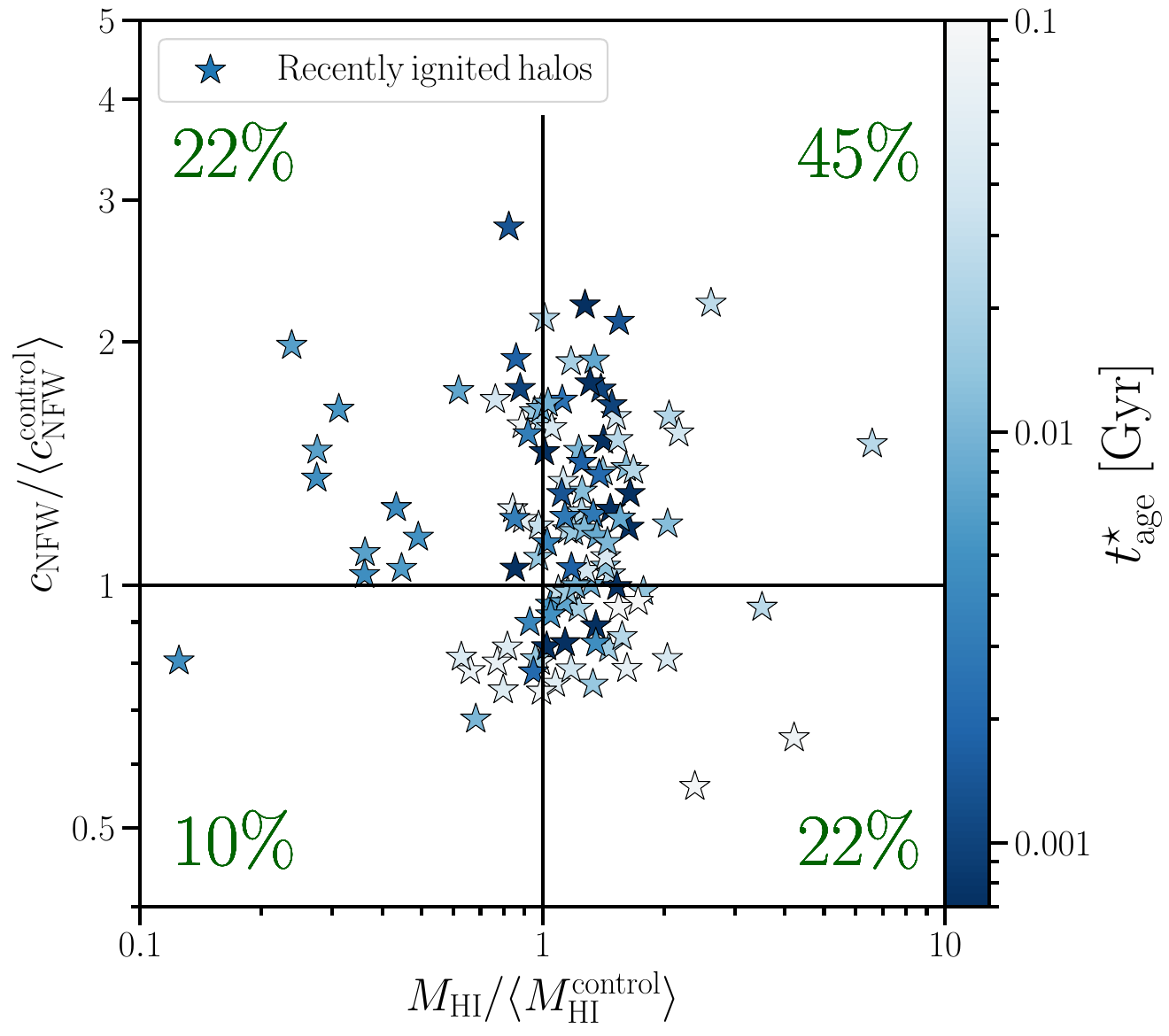}
    \includegraphics[width=0.49\textwidth]{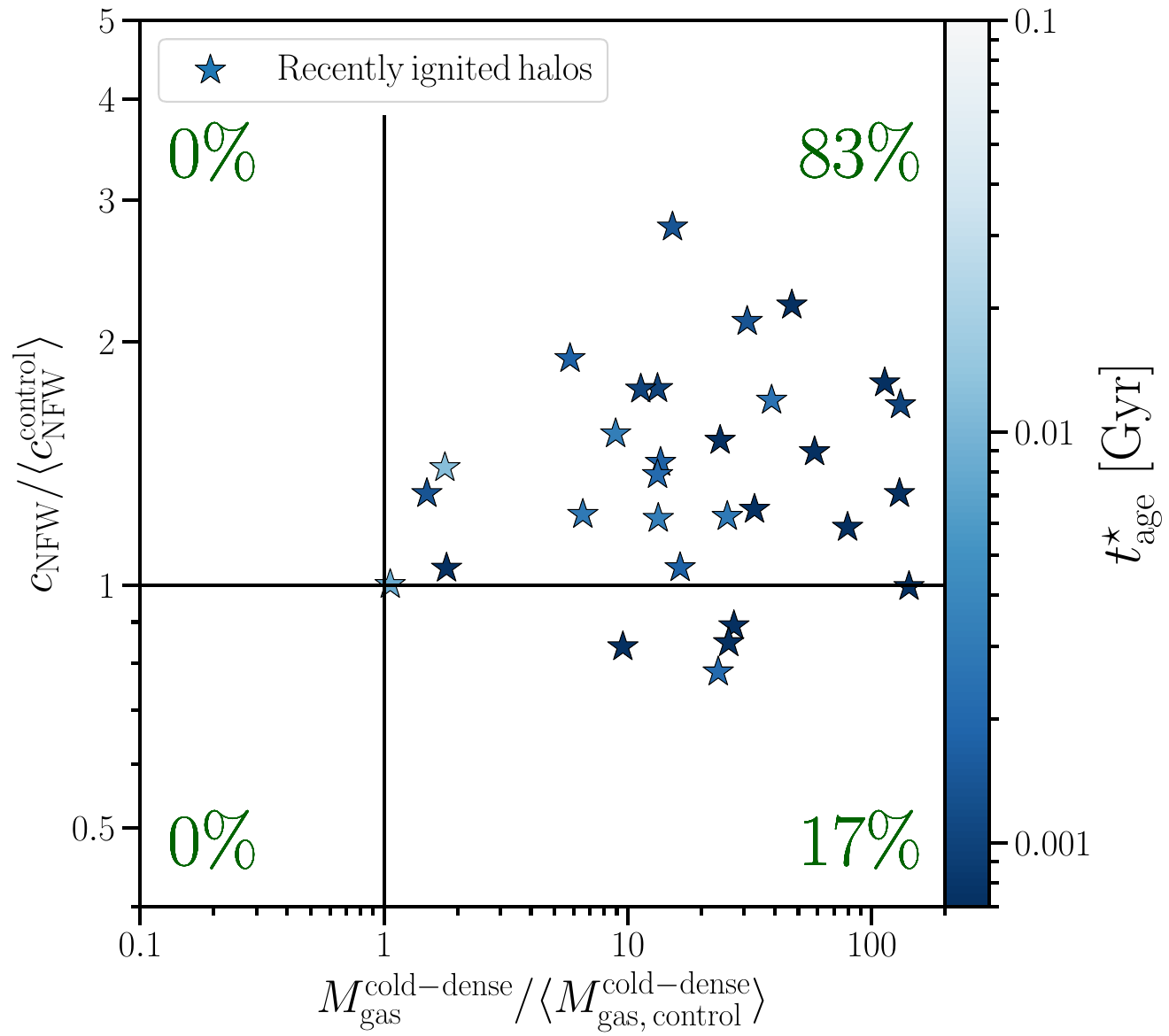}
    \caption{Enhancement of ISM and halo properties as drivers of galaxy ignition. NFW concentration enhancement versus HI gas mass enhancement ({\it left panel}) and cold-dense gas mass enhancement ({\it right panel}), depicted as { star-shaped symbols} and color-coded by stellar age of the recently-ignited halos. Here, enhancement is define as the ratio the property of a recently-ignited halo and the average value for across their matched failed-control counterparts. The vertical and horizontal solid lines denote unity. Approximately $68\%$ of our recently-ignited halos have either enhanced concentration or enhanced HI mass ($\sim45\%$ have both). In contrast, $100\%$ of recently-ignited halos have enhanced cold-dense gas budgets, $\sim83\%$ of which have also have enhanced halo concentrations. We note that these numbers refer exclusively to cases where both the recently-ignited halo and at least one of the matched fail halos have nonzero HI or cold-dense gas mass. We report these percentages per quadrant (large dark-green annotations).
    }
    \label{fig:delta_cnfw}
\end{figure*}

\subsection{Halo structure}\label{subsec:halo}

Recent numerical work suggests that halo properties, such as the depth of the potential, may play a role in the emergence of galactic disks and the transition for bursty to steady star formation in galaxies \citep{Hopkins2023,Benavides2025}. It is therefore reasonable to investigate if halo structure also influences the onset of galaxy formation. Beyond virial mass, we inspect the following properties: maximum circular velocity ($V_{\rm max}$), escape velocity ($V_{\rm esc}$), halo velocity dispersion ($\sigma_{v}$) and the Navarro-Frenk-White concentration \citep[$c_{\rm NFW}$,][]{NFW}. We find little connection between these quantities and galaxy-ignition, except for halo concentration.

Figure~\ref{fig:cnfw} explores possible connections between $c_{\rm NFW}$ and the mass of the following two ISM phases: HI (left panel) and cold-dense gas (right panel). The star-shaped symbols, color-coded by stellar age, represent the recently-ignited sample; while the gray squares denote the average values for each set of control failed halos matched by $M_{\rm vir}$, $M_{\rm gas}$ and $z$. The two populations exhibit strong overlap in terms of $M_{\rm HI}$. However, the concentrations of recently-ignited halos extends to much higher values.  In other words, there is a large subpopulation of recently-ignited halos with $c_{\rm NFW} \sim 4.5-10$ with (on average) no such failed-halo counterparts.

Cold-dense gas unveils a more striking story: these two populations are very segregated from each other on the concentration $-$ cold-dense gas mass plane. Namely, the overlap in on the concentration-HI-mass plane can be broken by inspecting cold-dense gas mass, where failed controls  tend to have $M_{\rm HI} \lesssim 2 \times 10^5 M_{\odot}$, and most recently-ignited objects do not. Nevertheless, strong overlap does not necessarily indicate that two populations are identical -- at least not on an object-by-object basis. 

Figure~\ref{fig:delta_cnfw} shows halo-concentration enhancements versus either HI-mass enhancements (left panel) or cold-dense gas mass enhancements (right panel), color coded by the ages of the targed recently-ignited stellar populations. There is no evident correlation between $c_{\rm NFW}$-enhancement and $M_{\rm HI}$-enhancement. We find that $\sim45\%$ of recently-ignited halos have both enhanced concentrations and enhanced HI-content. Likewise $\sim22\%$ have either enhanced concentrations and suppressed HI-content or suppressed concentrations and enhanced HI-content. Only $\sim10\%$ of objects have both suppressed concentrations and HI-content. In other words, even though Figures~\ref{fig:control_gas} and \ref{fig:cnfw} naïvely suggest no difference in HI-content between the two populations, we find that $\sim72\%$ of recently-ignited halos have either enhanced concentrations (by factors of up to $\sim3$) or enhanced HI-content (by factors of up to $\sim7$) relative to their matched failed counterparts. 

Replacing HI with cold-dense gas provides a more exaggerated version of this picture. First we find that $100\%$ of our objects exhibit enhanced cold-dense gas reservoirs, with enhancements exceeding factors of $100$. Likewise, we find that $83\%$ of this objects have enhanced halo concentrations. In summary we find that halo-concentration and cold-dense gas are reasonable properties to distinguish recently-ignited halos from failed halos with nearly identical virial mass and gas content at a given epoch. 
 

\section{Discussion}\label{sec:discussion}

\subsection{Galaxy ignition in other simulations}\label{threshold}

Using the same subgrid physics model as in the {\it Evolution and Assembly of GaLaxies and their Environments} project \citep[EAGLE,][]{Schaye2015,Crain2015}, \cite{Sawala2016} ran 12 zoom simulations aimed at reproducing the Local Group to investigate which low-mass halos host faint galaxies. Despite differences in subgrid physics and targeted volumes, these authors also find that several field halos in their simulations are ignited for the first time at $z\leq6$, in line with our results (their Figure 5 versus our Figure~\ref{fig:stellar_ages}). However, they report that such halos have masses between $\sim10^8-10^{10}M_\odot$, while we find a more restricted range ($\sim10^8-10^{9}M_\odot$; top-right panel of Figure~\ref{fig:evolution} and bottom panels of Figure~\ref{fig:stellar_ages}). It is possible that { a combination of differences in resolution} and their treatment of the interstellar medium and star formation is responsible for this \citep{DalaVecchia2008}. It would be interesting if that team revisited this question with their new {\it COLd Ism and Better REsolution} runs \citep[COLIBRE,][]{Schaye2025}, whose { resolution and} ISM-treatment are more similar to ours.

\cite{Sawala2016} also find that their luminous halos form earlier and have higher concentration than their dark counterparts. These authors use maximum circular velocity ($V_{\rm max})$ to quantify halo structure, and claim that objects with higher $V_{\rm max}-$values are expected to cool gas more efficiently and inhibit reionization-driven photoevaporation. Their Figure~7 shows that $V_{\rm max}$ for their luminous field halos is slightly higher than that of their non-luminous counterparts. However, this enhancement is below $\sim$0.5 dex, and unfortunately they do not quantify scatter, making it impossible to judge the significance of this deviation. That figure also shows that these two populations occupy very distinct $M_{\rm peak}$ regimes, with only a narrow overlap between $\sim2-6 \times 10^9 M_\odot$. By matching by $M_{\rm vir}$ and $M_{\rm gas}$, we find that our recently-ignited and failed-control halos have very similar maximum circular velocities (within scatter, not shown). Nevertheless, the central finding of our work is that galaxy ignition is connected to concentration ($c_{\rm NFW}$, Figures~\ref{fig:cnfw} and \ref{fig:delta_cnfw}), underscoring the importance of halo structure.

As a successor of \cite{Sawala2016}, the {\it A Project Of Simulating The Local Environment} runs \citep[APOSTLE,][]{Fattahi2016,Sawala2016b} simulate 12 Local-Group analogs at three resolution levels. Using this project, \cite{Benitez-Llambay2017} determine that failed halos come in two varieties: those suppressed by reionization and those ram-pressure stripped by the cosmic web \citep{Benitez-Llambay2013}. Given our interest in galaxy ignition -- in particular the role of the ISM -- here we focus primarily on halos with a resolved gas component (with at least 128 gas particles). We plan to study gas-devoid failed (and ignited) halos, and the role of the cosmic web, in future work. 

In a follow up paper, \cite{Benitez-Llambay2021} investigate the tail of late-forming low-mass galaxies. Their Figure~2 suggests that $8\%$ of their halos host luminous galaxies at $z<3$, in stark contrast with our results. Indeed, the last halo to ignite galaxy formation in our simulation appears at $z=2$ (Figure~\ref{fig:awake_vs_failed}). This difference with our work could be driven by their more relaxed threshold for star formation. Another possibility is that the Local Group environment in their simulations is instigating galaxy formation, akin to more massive satellites experiencing gas compression in the outskirts of groups and clusters \citep{Bekki2014}. Interestingly, in a more recent paper \citep{Herzog2023}, that team conducts a similar study beyond the Local Group environment. Concretely, they utilize a cosmological simulation similar to EAGLE, but with higher resolution and smaller volume \citep{Benitez-Llambay2020}. It would be interesting to see if isolated halos in that simulation also form galaxies at $z<3$.

Using the IllustrisTNG simulation \citep{Pillepich2018,TNG50}, \cite{Lee2024}  investigate the formation and evolution of dark galaxies, which they define to have $M_\star/M_{\rm total} < 10^{-4}$. These authors elect to constrain their `star-poor' and `starless' samples to have $M_{\rm dm} = 1-3 \times 10^9 h^{-1} M_\odot$ (i.e. $M_{\rm dm} = 1.4-4.3 \times 10^9 M_\odot$) at $z=0$ to ensure a fair comparison (pink band in their Figure~2). The upper right panel of our Figure~\ref{fig:evolution} displays a similar overlap at $z=0$ between our failed and ignited populations ($M_{\rm vir} = 2.5-3.5 \times 10^9 M_\odot$). However, we also show that the range of masses covered by these populations exhibits a strong evolution, about an order of magnitude increase between $z=6$ and today. Thus, the fact that these authors chose to anchor their populations to $z=0$ hinders a comparison between their work and ours across epochs (i.e., their Figure~5 versus our Figure~\ref{fig:evolution}). 

Figure~3 of \cite{Lee2024} shows that, at $z<6$, both their star-poor and star-less halos tend to have gas with temperatures $\sim10^{4}-10^5$ K. In other words, for both populations, gas is primarily in their designated `halo gas' phase \citep{Martizzi2019}. These populations also exhibit a small fraction of star-forming gas (with $n_{\rm H}>0.1$ cm$^{-3}$), which diminishes with cosmic time. Namely, despite their cruder gas treatment\footnote{The IllustrisTNG subgrid model does not fully resolve the ISM, and instead imposes a temperature floor at $T=10^4$ K \citep{SH03}.}, this work also suggests that galaxy-ignition peters out at late epochs (our Figures~\ref{fig:cosmic_web}, \ref{fig:mass_function} and \ref{fig:stellar_ages} -- see also their Figure~4). One major difference is that none of their star-poor and starless halos contain hot gas (with $T>10^6$ K). The bottom-right panel of our Figure~\ref{fig:control_gas} shows that while only two objects in our recently-ignited sample contain hot gas, the presence of this gas phase is ubiquitous among failed halos. We suspect that this is caused by our more detailed treatment of feedback physics. We defer a more extensive analysis on the ISM properties of failed halos to future work.

The bottom panels of Figure~5 in \cite{Lee2024} also show that their star-poor halos have slightly higher densities within their half-mass radii than do their starless counterparts -- in line with our findings on the role of halo concentration and dense gas in galaxy ignition -- though their large $1\sigma$ scatter precludes us from drawing firm conclusions. It would be interesting if these authors conducted a matched-control comparison akin to our Figures~\ref{fig:cnfw} and \ref{fig:delta_cnfw}.

On a similar vein, \cite{Jeon2025} investigate the nature of starless subhalos in the vicinity\footnote{Within $1.5\times R_{\rm 200,c}$ of these Milky Way analogs, where $R_{\rm 200,c}$ is defined as the radius within which the enclosed density is 200 times the critical density} of 26 Milky Way analogs selected from the N{\small EW}H{\small ORIZON} and N{\small EW}H{\small ORIZON2} simulation volumes \citep{Dubois2021,Yi2024}. Similar to the approach by \cite{Lee2024}, these authors also concentrate on the mass range where the starless and starred populations overlap (in this case, $M_{\rm vir} \sim 10^{8.4}-10^{8.9} M_\odot$). They find that these two populations have similar amounts of gas. We also find that failed and ignited halos have similar gas content (middle-left panel of our Figure~\ref{fig:evolution}), but our ignited halos are centrals with a single star particle. One key difference between their samples and ours is that their starless population contains no cold gas (defined to have $T<10^4$ K), in stark contrast with our failed halos, many of which contain cold-dense\footnote{Recall that we define cold-dense gas to have temperature $< 10^2$ K and density $> 10^2$ cm$^{-3}$ \citep{Moreno2019}.} gas (bottom-left panel of our Figure~\ref{fig:control_gas}). Indeed, some of our failed halos even contain star-forming gas (top-right panel of our Figure~\ref{fig:control_gas}). This could be attributed to differences in subgrid physics, including their more relaxed gas density threshold for star formation. Despite this, unlike the other simulations described above (and like ours), this model also captures the multi-phase structure of the ISM. We suspect that this difference is driven rather by the fact that these authors selected objects in close proximity to Milky Way analogs. We defer investigations on the role of environment to future work.

We close this section by underscoring the importance of properly comparing ignited halos to their failed counterparts. As a first step, both \cite{Lee2024} and \cite{Jeon2025} constrain their samples to overlapping mass scales to ensure a fair comparison. While we acknowledge that such a step is important, we argue that this does not suffice. To illustrate, the left panel of Figure~\ref{fig:control_gas} might naïvely suggest that our recently-ignited and failed-match halo populations have identical HI properties. However, the left panel of Figure~\ref{fig:delta_cnfw} demonstrates that approximately two thirds of recently-igniting halos have enhanced HI-mass content, and a non-negligible fraction experiences above $100\%-$level enhancements.

\subsection{The dark halo occupation fraction}\label{subsec:threshold}

Recent years have witnessed a boom of analytic, semi-empirical and semi-analytic / forward-modelling studies focused at constraining the faint end of the galaxy-halo connection, and seeking to unveil the minimum halo mass that can host a galaxy -- often quantified by the halo occupation fraction \citep{Benitez-Llambay2020,Manwadkar2022,Nadler2024,Ahvazi2024,Nadler2025}. In tandem, several cosmological simulations with various levels of sophistication have also attempted to tackle these questions \citep{Okamoto2008,Sawala2016,Fitts2017,Benitez-Llambay2017,Jethwa2018,Benitez-Llambay2020,Munshi2021,Kim2024,Doppel2025}. However, \cite{Munshi2021} rightfully warns against this practice because the shape and location of the halo occupation function depends on resolution (see their Figure~4). We agree with this indictment, and emphasize that our results below should be interpreted as the fraction of halos unable to form galaxies with $M_\star>6.26\times10^4M_\odot$ (corresponding to a single stellar particle at our chosen resolution) under the subgrid assumptions adopted by our \texttt{FIRE-2} physics model. 

In other words, our goal here is not so much to reach the `true' threshold of galaxy formation, but to test our \texttt{FIRE-2} physics model using one of the large-volume cosmological simulations with the highest dynamical range in existence today \citep[\texttt{FIREbox},][]{Feldmann2022}. Comparisons against our results by others are straightforward: by simply constraining their predictions to halos above our minimum stellar mass. Indeed, it is common practice to sacrifice resolution for volume (and vice versa). In our case, it is certainly reassuring that our principal findings agree with \cite{Fitts2017}. Namely, that low concentration might be the culprit behind halos failing to form stars -- despite the fact their small-volume zoom simulations have over $100$ times higher baryonic mass resolution (but the same \texttt{FIRE-2} physics model). The advantage here is that we are able to extend this result to larger, cosmologically-meaningful volumes -- and bring attention to the second (perhaps more important) missing ingredient: the presence of cold-dense gas.

\begin{figure}
    \centering
    \includegraphics[width=0.45\textwidth]{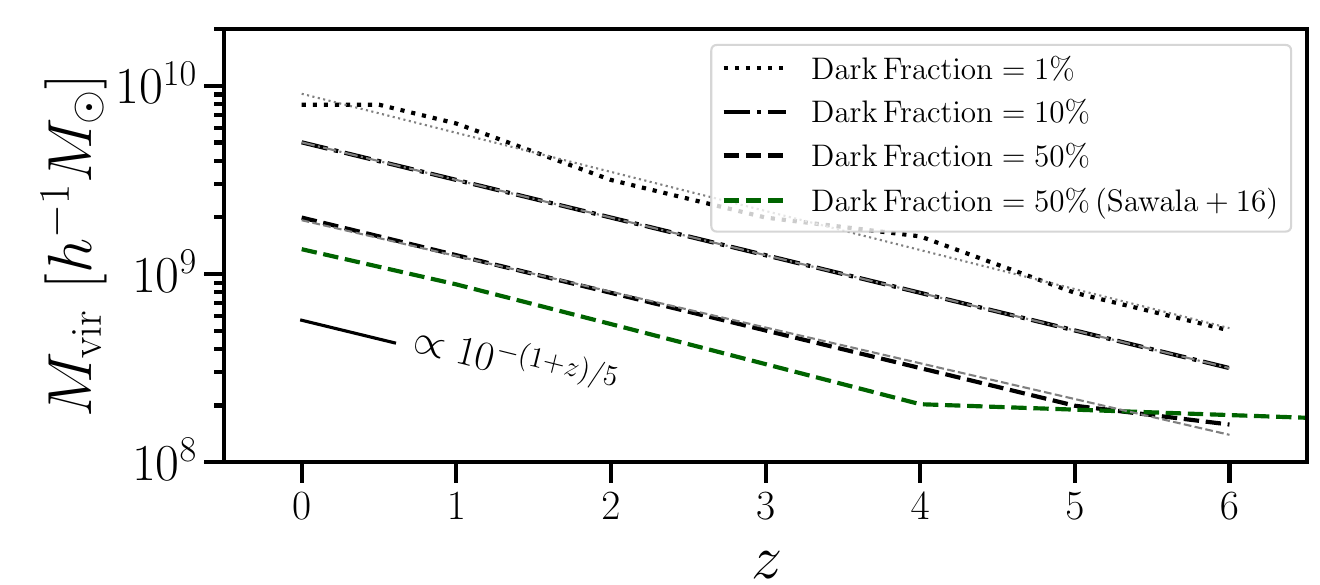}
    \includegraphics[width=0.45\textwidth]{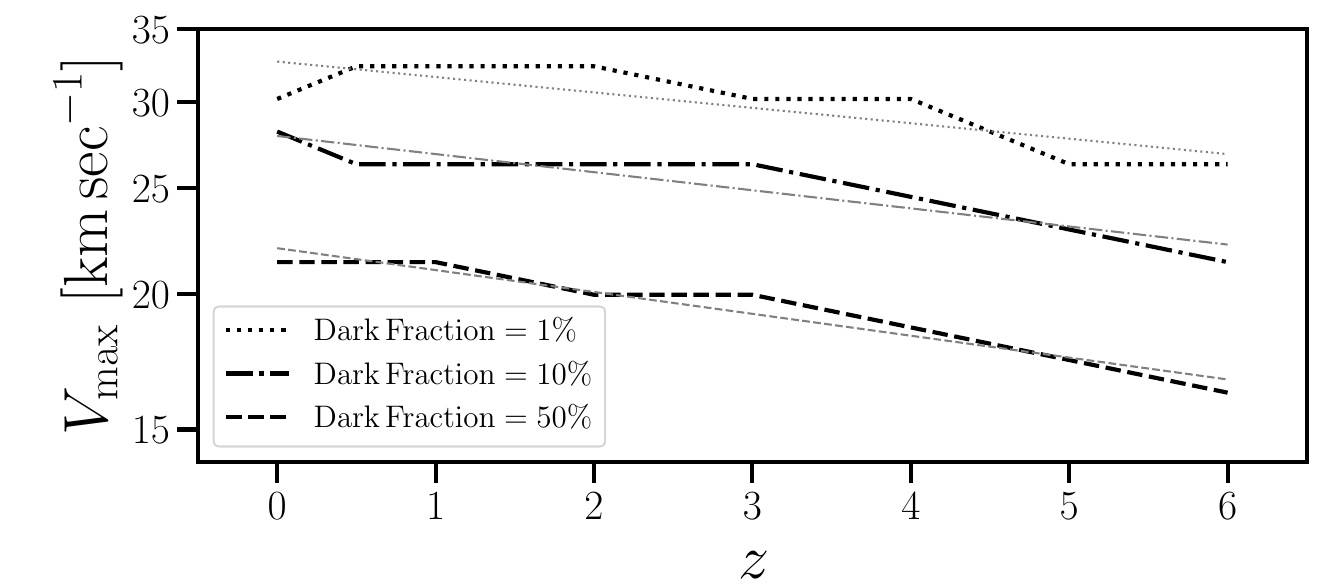}
    \caption{The evolution of the dark fraction, defined as the comoving number density of all failed halos divided by that of all halos (Figure~\ref{fig:mass_function} and Table~\ref{table:samples}). Recall that we exclude subhalos in our analysis. {\it Top panel:} The value of $M_{\rm vir}$ at which $x\%$ of all halos are failed, with $x=1$, $10$ and $50$ (dotted, dot-dashed, dashed and solid black lines). These values diminish with $z$ as $\sim10^{-(1+z)/5}$ (slanted black segment indicates this behavior). The thin gray lines denote linear best fits. { We also include $x=50$ values from \citep{Sawala2016} for comparison (green dashed line).} {\it Bottom panel:} Same as above, but in terms of maximum circular velocity $V_{\rm max}$. The redshift dependence is shallower in this case.
    }
    \label{fig:thresholds}
\end{figure}

Figure~\ref{fig:thresholds} investigates the threshold at which $x$ percentage of halos are dark in our simulation. Recall that the bottom panels of Figure~\ref{fig:mass_function} report halo fractions, defined as the ratio of the halo mass function for a given population modulo that of the `all halos' population. Recall also that we exclude subhalos from our analysis. The dark fraction here is therefore defined as the halo fraction corresponding to the `all-failed halo' population -- i.e., the largest $M_{\rm vir}$ values at which the black (and gray) curves intersect the horizontal dotted, dot-dashed and dashed black lines. 

To illustrate, Figure~\ref{fig:mass_function} shows that at $z=0$, { over half of all halos with $M_{\rm vir} \sim 2 \times 10^9 h^{-1}M_\odot$ are devoid of stars. This value shifts to $\sim 5 \times 10^8 h^{-1}M_\odot$ and $\sim 1.5 \times 10^8 h^{-1}M_\odot$ at $z=3$ and $z=6$.} The top panel of Figure~\ref{fig:thresholds} connects these values across every target redshift considered in this study. The dotted, dot-dashed and dashed black lines represent the value of $M_{\rm vir}$ as a function of $z$ at which the dark fraction equals $1\%$, $10\%$ and $50\%$ respectively. These thresholds evolve approximately as 
\begin{equation}
  M_{\rm vir}(z) \propto 10^{-(1+z)/5},
\end{equation}\label{eqn:df}
as indicated by the inclined black solid segment. We note that using a very different model, \cite{Sawala2016} also find that this dark halo fraction evolves with redshift, with similar halo mass values at the $50\%$ level\footnote{It is difficult to compare the $10\%$ and $1\%$ levels directly against this and other works because the $y-$axis is often presented using a linear scale.} { (within a factor of $\sim2$, as indicated by the green dashed lines -- see also their Figure 3).}

We also perform best-linear fits (dotted, dot-dashed and dashed thin gray lines) and find that this threshold evolution can be characterized as
\begin{equation}
    \log \,[M^{x}_{\rm vir}(z)/h^{-1}M_{\odot}] = 9+\log A_{x} - (1+z)/\alpha_{x},
\end{equation} \label{eqn:df_fits}
where $(A_{1},A_{10},A_{50}) = (15,8,3)$ and $(\alpha_{1},\alpha_{10},\alpha_{50}) = (4.8,5,5.3)$. 

The bottom panels present a similar analysis, but now in terms of halo maximum circular velocities. Many authors prefer to characterize the onset of galaxy formation in terms of this quantity \citep{Rees1986,ThoulWeinberg1996,Okamoto2008uv} because it is more closely aligned to the dynamical mass of the halo at the time of collapse. In other words, it is more resilient than total mass, (1) which increases as the halo accretes satellites/subhalos plus diffuse matter \citep{Zhao2003,Li2007,Zehavi2019}, and (2) decreases due to environmental processes if said halo becomes a subunit within a larger system \citep{Kravtsov2004,Diemand2007,Kuhlen2007,DiCintio2011}. 

In our simulation we find a shallower redshift evolution, which can be characterized as 
\begin{equation}
    \log \, [V^{x}_{\rm max}(z)/{\rm km \sec}^{-1}] =\log B_{x} - (1+z)/\beta_{x},
\end{equation} \label{eqn:df_fits2}
where $(B_{1},B_{10},B_{50}) = (34,29,23)$ and $(\beta_{1},\beta_{10},\beta_{50}) = (70,60,50)$. It would be interesting to see how Figure~3 in \cite{Sawala2016} and Figure~2 in \cite{Doppel2025} would change if they were to express their results in terms of maximum circular velocity -- and if the semi-empirical / semi-analytic models cited above produce a similarly strong mass evolution and weaker $V_{\rm max}$ trends as in our Figure~\ref{fig:thresholds}.

\section{Conclusions}\label{sec:conclusions}

We use \texttt{FIREbox} \citep{Feldmann2022}, a large-volume cosmological simulation employing the Feedback In Realistic  Environments physics model \citep[\texttt{FIRE-2},][]{FIRE2}, to identify the physical processes responsible for igniting galaxy formation in halos in the post-reionization era.

Below we summarize our findings, followed by remarks on caveats, limitations and possible future directions.

\begin{itemize}
    \item Although smaller in number, the ignited halo population  exhibits an evolution similar to the failed halo population -- in terms of virial mass, total gas mass (all ISM phases included) and HI gas. 
    \item However, recently-ignited halos (with stellar ages below $100$ Myr) evolve differently. Although they do exist in the late universe ($2 \leq z \leq 6$), their abundance diminishes dramatically with cosmic time. These objects often contain larger gas and HI reservoirs, as well as higher $M_{\rm gas}/M_{\rm vir}$ and $M_{\rm HI}/M_{\rm gas}$ ratios.
    \item The recently-ignited halo population and their carefully-matched failed control sample (within $0.1$ dex in $M_{\rm vir}$ and $M_{\rm gas}$, and at the same $z$)  overlap in terms of HI content, but are segregated from each other when other ISM phases are considered.
    \item This segregation is $\sim 2$ dex in star-forming gas mass, but only $\sim9\%$ of recently ignited halos contain this phase (and only $\sim28\%$ of failed control sets include at least one halo with this type of gas).
    \item Similarly, segregation is $\sim 2-3$ dex in cold-dense gas mass \citep[following][]{Moreno2019}, but only $\sim21\%$ of recently ignited halos contain this type of gas ($\sim88\%$ of failed control sets include at least one halo with this ISM phase).
    \item Only $\sim1\%$ of recently-ignited halos harbor hot gas (with temperatures $>1$ million K). Approximately $70\%$ of the matched-control sets contain at least one failed halo with hot gas -- but on average, these samples have lower hot gas mass content.
    \item The recently-ignited halo population and its matched failed control counterpart exhibit some overlap on the $c_{\rm NFW}-M_{\rm HI}$ plane, but segregate strongly on the $c_{\rm NFW}-$ cold-dense gas mass plane. Concretely, recently-ignited halos tend to have richer cold-dense gas reservoirs and some tend to have much higher NFW halo concentrations.
    \item On an object-by-object basis, $\sim45\%$ of the recently-ignited halos exhibit enhancements in both $c_{\rm NFW}$ and $M_{\rm HI}$, by factors up to $\sim3$ and $\sim10$ respectively. On the other hand, $\sim22\%$ display enhancements in one of these quantities and suppression on the other. Approximately $10\%$ are suppressed in both $c_{\rm NFW}$ and $M_{\rm HI}$.
    \item Similarly, $100\%$ of recently-ignited halos experience cold-dense gas mass enhancements, by factors up to $\sim100$, relative to their matched failed counterparts. Approximately $83\%$ exhibit both cold-dense gas mass and $c_{\rm NFW}$ enhancements, while the remaining $\sim17\%$ have suppressed NFW halo concentrations (but still enhanced cold-dense gas mass content).

\end{itemize}

Our results align with an emerging picture in which gas compression is more efficient in more concentrated halos, thus promoting gas cooling and, ultimately, galaxy ignition. These high concentrations might be governed by formation time, spin, large-scale environment, assembly history, recent merger events { (e.g., a soon-to-be descendant of the merging failed halo in Figure~\ref{fig:awake_vs_failed})} -- or a complex combination of these physical processes \citep{Wechsler2002,Giocoli2012,Ludlow2014,SatoPolito2019,Wang2020,Hellwing2021}. Once the galaxy is ignited, new feedback processes materialize under non-trivial timescales, which may further treat the ISM in unexpected ways \citep{Stilp2013,Hunter2025} -- i.e., by injecting heat / turbulence or rarefying / depleting the gas reservoir -- thus further complicating the usage of current ISM properties as a signpost for recent galaxy ignition. Addressing this complexity is the subject of future work by us and (hopefully) others. Meanwhile, we underscore the importance of large-volume simulation studies like ours, which quantify the frequency of galaxy ignition and its dependence on other properties -- such as mass, epoch, ISM content, halo structure, and so forth -- in a cosmologically meaningful fashion.

Future work with our simulations also includes an exploration into the early universe, prior and during the epochs of reionization. To achieve this, we expect to use \texttt{BonFIRE}, a cosmological simulation similar to \texttt{FIREbox}, but with higher resolution and a larger volume (Samuel et al., in prep). We also plan to undertake a more thorough characterization of failed halos with a resolved gas component. Figure~\ref{fig:control_gas} suggests that this population contains HI, while subsets of this sample are endowed with hot, cold-dense and even star-forming gas -- thus making them potentially observable \citep{Leisman2021,Xu2023,Anand2025,Kwon2025,Montes2025,Sun2025}. Furthermore, in this paper we focus exclusively on centrals to avoid the effect of environment. However, even centrals experience the large scale tidal field \citep{Mercado2025} and ram-pressure stripping from the cosmic web \citep{Benitez-Llambay2013}. It is possible that location on the cosmic web influences a halo's ability to accrete \citep[as suggested by ][]{Lee2024,Jeon2025} and retain gas. It would also be interesting to determine if Milky Way / Local Group environments are likely to attract these class of halos, enabling the detection of their descendants in the foreseable future. We plan to investigates connections between galaxy-ignition (or failure) and the large-scale density / tidal field, plus the proximity to Milky Way  / Local Group analogs (Figure~\ref{fig:cosmic_web}) in future papers. 

Albeit its immense dynamical range, the \texttt{FIREbox} cosmological simulation, like any other numerical treatment, is not free of caveats and limitations. As pointed out by \cite{Munshi2021}, resolution limits our ability to probe the onset of galaxy formation. To address this, our collaboration is running \texttt{WildFIRE}, a massive suite of zoom simulations (see also Wheeler et al., in prep, for similar results). Now, there is a limit to how much one can benefit from increased resolution, which is governed by the fact that different physical processes must be invoked in that regime, especially as we approach the resolved single-star limit \citep{Lahen2020,Andersson2023,Deng2024,Brauer2025}. { For this reason, we plan to simulate a handful of \texttt{WildFIRE} runs with {\it STAR FORmation in Gaseous Environments} (\texttt{STARFORGE}) physics \citep{STARFORGE}, using the hybrid FORGE'd in FIRE setup similar to that discussed in \cite{FORGED}.}  

Addressing small-scale challenges certainly alleviates physical uncertainties, but limitations at large scales remain. Concretely, our simulations assume a homogenous, spatially uniform UV background and lacks the incorporation of radiative transfer processes. Indeed, reionization is a patchy process that affects different regions of the universe at different times \citep{Renaissance,SPHINX,CODA,THESAN}. Unfortunately, self-consistent radiation-hydrodynamic simulations (at the resolution and volume attained by \texttt{FIREbox}) are still prohibitively expensive  -- although a much cheaper alternative could include the incorporation of excursion sets \citep{Furlanetto2004,Majumdar2014}. 

Lastly, galaxy ignition within the smallest eligible halos in the universe may be impacted by the very nature of dark matter itself. Indeed, simulations show that replacing cold with warm or self-interacting dark matter (WDM or SIDM) affects halo structure and abundance \citep[see e.g.,][and references therein]{BullockMBK2017}. For instance, WDM is believed to suppress structure formation at small scales, limiting the number of halos elegible for galaxy ignition \citep{Angulo2013,Lovell2014}.
Likewise, SIDM halos tend to be less centrally concentrated than their CDM counterparts \citep{Vogelsberger2014,Fitts2019}, potentially subverting galaxy ignition.  

Ultimately, the real Universe is the final judge. For this reason, the query of gas-rich ultra-faint galaxies with young stellar populations beyond the Local Group is imperative. The recent discoveries of isolated gas-rich star-forming low-mass galaxies such as Leo P, Pavo, Corvus A, and Kamino, and their star formation histories \citep{Giovanelli2013, McQuinn2024, Jones2024, Mutlu-Pakdil2025} highlight mass growth in some of the smallest systems in the post-reionization era. Ideally, analogs to these systems – but fainter and lacking an ancient stellar population (i.e. no star formation during or before the Epochs of Reinization) – might shed light on galaxy ignition (and failure) in the post-reionization universe.  
For this, we might need to wait a few decades until the next generation of Extremely Large Telescopes \citep[ELTs,][]{Macri2024} and forthcoming large radio astronomy projects -- such as the {\it Next Generation Very Large Array} \citep[ngVLA,][]{ngVLA} become available. Another promising avenue is gravitational lensing \citep{Vegetti2014,Despali2025,Lange2025}, which may potentially allow us to catch the process of galaxy ignition `in the act' in the not-so-distant future. Meanwhile, comprehensive simulation work like ours is essential to guide this imminent golden age of faint extragalactic astronomy.


\begin{acknowledgments}
JM dedicates this paper to his beloved late father, Juan Moreno P\'erez (RIP), on his journey back to the stars. JM is funded by a Pomona College Large Research Grant and thanks the Carnegie Observatories for their hospitality. CW acknowledges the CPP Provost's Teacher-Scholar Program. FJM is funded by the National Science Foundation (NSF)
Math and Physical Sciences (MPS) Award AST-2316748. MKRW acknowledges support from NSF MPS Ascending Faculty Catalyst Award AST-2444751. JS and MBK acknowledge support for program JWST-AR-06278 by NASA through a grant from the Space Telescope Science Institute, which is operated by the Association of Universities for Research in Astronomy, Inc., under NASA contract NAS 5-03127. EC and RF acknowledge financial support from the Swiss National Science Foundation (grant no 200021-188552). MBK also acknowledges support from NSF grants AST-1910346, AST-2108962, and AST-2408247; NASA grant 80NSSC22K0827; HST-GO-16686, HST-AR-17028, HST-AR-17043, and JWST-GO-03788, from STScI; and from the Samuel T. and Fern Yanagisawa Regents Professorship in Astronomy at UT Austin. AW re-ceived support from NSF, via CAREER award AST-2045928 and grant AST-2107772. Support for PFH was provided by a Simons Investigator Award. { The authors wish to thank Alejandro Benítez Llambay, Martin Rey and Ethan Nadler for thoughtful suggestions on an earlier draft, and the anonymous reviewer for their insightful comments.} The authors also acknowledge the labor by the cleaning and the clerical staff, the food service workers, the technical support personnel, and many others that make the astronomy research enterprise possible. This work was conducted on Tongva-Gabrielino land.\\
\end{acknowledgments}

\begin{contribution}
JM wrote the entire manuscript and led every aspect of the project. CW, FJM, and MKRW made significant contributions to the original design of the paper. MKRW also provided unique observational insight. JS and PJG provided special support regarding connections to the early universe, while EC and RF created the simulations and AHF catalogs. MBK, AW, JBS and PFH provided additional support on the big-picture context of the project and the creation of our physics model. All authors contributed substantially to the final polishing of this manuscript.

\end{contribution}

\software{yt \citep{Turk2011}, Amiga Halo Finder \citep{Knollmann2009}.}

\bibliography{bibliography}{}
\bibliographystyle{aasjournalv7}

\end{document}